\documentclass[12pt,a4paper,onecolumn,draftcls]{IEEEtran}
\usepackage{epsfig,graphicx,subfigure,psfrag,amsmath,cases}
\usepackage{latexsym,amssymb,epsfig,algorithm}
\usepackage{algorithmic}
\usepackage[usenames, dvipsnames]{color}
\usepackage{url}
\usepackage{scrtime}
\usepackage{bm}
\usepackage{framed}
\usepackage{epstopdf}
\usepackage{caption}
\author{{Elena Boshkovska, Derrick Wing Kwan Ng, Nikola Zlatanov, Alexander Koelpin,
and Robert Schober} \thanks{This paper has been presented in part at IEEE ICC 2016 \cite{ICC_non_linear} and at SPAWC 2016 \cite{SPAWC_non_linear}. Elena Boshkovska, Alexander Koelpin, and Robert Schober are with the Friedrich-Alexander-University Erlangen-N\"urnberg (FAU), Germany. Derrick Wing Kwan Ng is with The University of New South Wales, Australia.  Nikola Zlatanov is with the Monash University, Australia.}}

\title{Robust Resource Allocation for MIMO Wireless Powered Communication Networks Based on a Non-linear EH Model}

\date{}

\newtheorem{Thm}{Theorem}
\newtheorem{Lem}{Lemma}

\newtheorem{T-Prob}{Transformed Problem}
\newtheorem{proposition}{Proposition}
\DeclareMathOperator{\Tr}{\mathrm{Tr}}
\DeclareMathOperator{\zero}{\mathbf{0}}
\DeclareMathOperator{\Rank}{\mathrm{Rank}}

\DeclareMathOperator{\maxo}{\mathrm{maximize}}

 \newcommand{\qed}{\hfill \ensuremath{\blacksquare}}

\newcommand{\norm}[1]{\lVert#1\rVert}
\allowdisplaybreaks

\begin{document}

\maketitle
\vspace*{-1.5cm}
\begin{abstract}

In this paper, we consider a multiple-input multiple-output wireless powered communication network (MIMO-WPCN), where multiple users harvest energy from a dedicated power station in order to be able to transmit their information signals to an information receiving station. Employing a practical non-linear energy harvesting (EH) model, we propose a joint time allocation and power control scheme, which takes into account the uncertainty regarding the channel state information (CSI) and provides robustness against imperfect CSI knowledge. In particular, we formulate two non-convex optimization problems for different objectives, namely system sum throughput maximization and maximization of the minimum individual throughput across all wireless powered users. To overcome the non-convexity, we apply several transformations along with a one-dimensional search to obtain an efficient resource allocation algorithm. Numerical results reveal that a significant performance gain can be achieved when the resource allocation is designed based on the adopted non-linear EH model instead of the conventional linear EH model. Besides, unlike a non-robust baseline scheme designed for perfect CSI, the proposed resource allocation schemes are shown to be robust against imperfect CSI knowledge.
\end{abstract}
\vspace*{-3mm}

\newpage
\section{Introduction}
In recent years, wireless energy transfer (WET) has attracted a significant amount of attention in both academia and industry as a sustainable approach for supplying energy to low-power wireless communication devices, such as wireless sensors \cite{JR:Xiaoming_magazine}--\nocite{Ding2014,Krikidis2014,CN:WIPT_fundamental,JR:WIP_receiver,JR:MIMO_WIPT,
JR:EE_SWIPT_Massive_MIMO,JR:Kwan_secure_imperfect,JR:MOOP,JR:OFDM_relay_SWIPT,Powercast,JR:WPC_Rui_Zhang,JR:WPC_QQ,CN:User_Coop_WPCN,JR:WPCN_full_duplex}\cite{CP:optimal_time_all_WPCN_globecom_2014}.  With WET technology, energy-limited wireless devices can harvest energy from their received radio frequency (RF) signals to recharge their batteries and prolong their lifetimes. In fact, RF signals offer a more controllable and relatively stable energy source compared to the natural renewable sources available for energy harvesting (EH), such as solar and wind \cite{CN:WIPT_fundamental}. Additionally, RF signals can serve as a dual purpose vehicle for transporting both information and energy signals via the same carrier, which facilitates simultaneous wireless information and power transfer (SWIPT).

Besides SWIPT, another emerging line of research considers WET for wireless powered communication networks (WPCNs), where wireless communication devices first harvest energy, either from a dedicated power station or from ambient RF signals, and then use the harvested energy to transmit information signals \cite{JR:WPC_Rui_Zhang,JR:WPC_QQ}. Over the past few years, resource allocation algorithm design for SWIPT systems \cite{JR:MIMO_WIPT}--\nocite{JR:EE_SWIPT_Massive_MIMO,JR:Kwan_secure_imperfect,JR:MOOP}\cite{JR:OFDM_relay_SWIPT} and WPCNs \cite{JR:WPC_Rui_Zhang}--\nocite{JR:WPC_QQ,CN:User_Coop_WPCN,JR:WPCN_full_duplex,CP:optimal_time_all_WPCN_globecom_2014}\cite{JR:sum_rate_mimo_WPCN_rank_one} has been extensively studied. However, the most critical challenge in supplying a sufficient amount of energy efficiently for wireless devices in the far-field via WET still persists.  In particular, wireless power has to be transferred
via a carrier signal with a high carrier frequency such that antennas with reasonable size can be used for harvesting the power. Thus, with increasing distance between the wireless devices and the wireless power supply station, the propagation path loss attenuating the signal during WET also increases significantly \cite{JR:far_field_wpt}. A viable approach for increasing the amount of harvested energy is to improve the efficiency of the RF EH circuits employed by the wireless devices to convert the collected RF energy to electrical energy. To this end, a considerable amount of work has been devoted to the optimization of practical RF EH circuits, employing various hardware architectures \cite{JR:Energy_harvesting_circuit}\nocite{JR:EH_measurement_1}--\cite{CN:EH_measurement_2}. On the other hand, the design of efficient resource allocation schemes in WET systems relies on accurate mathematical models for the adopted RF EH circuit. Unfortunately, most of the existing works, in both the SWIPT and the WPCN literature assume an overly simplistic linear EH model for characterization of the RF energy-to-direct current (DC) power conversion since the resulting resource allocation problems are relatively easy to solve \cite{CN:WIPT_fundamental}--\cite{JR:OFDM_relay_SWIPT}, \cite{JR:WPC_Rui_Zhang}--\cite{JR:far_field_wpt}. However, in practice, the conversion efficiency is a fundamental performance metric for RF EH circuits, and various experiments for practical EH circuits have shown that their input-output characteristic is highly non-linear \cite{JR:Energy_harvesting_circuit}\nocite{JR:EH_measurement_1}--\cite{CN:EH_measurement_2}.  The discrepancy between the properties of practical non-linear EH circuits and the linear EH model conventionally assumed in the SWIPT \cite{JR:MIMO_WIPT}--\nocite{JR:EE_SWIPT_Massive_MIMO,JR:Kwan_secure_imperfect,JR:MOOP}\cite{JR:OFDM_relay_SWIPT} and WPCN \cite{JR:WPC_Rui_Zhang}--\nocite{JR:WPC_QQ,CN:User_Coop_WPCN,CP:optimal_time_all_WPCN_globecom_2014}\cite{JR:sum_rate_mimo_WPCN_rank_one} literature may cause severe resource allocation mismatches, leading to significant performance degradation in practical implementations.

Recently, the authors of \cite{Letter_non_linear} proposed a practical parametric non-linear EH model to capture the non-linear characteristics of the end-to-end WET. The non-linear EH model proposed in \cite{Letter_non_linear} was exploited for the design of a beamforming algorithm for a downlink multiple antenna SWIPT system serving multiple information receivers  and multiple EH receivers. It was shown in \cite{Letter_non_linear} that resource allocation schemes designed based on this non-linear EH model yield a significantly higher amount of harvested energy compared to those designed for the traditional linear EH model. In \cite{ICC_non_linear} and \cite{SPAWC_non_linear}, the non-linear EH model was further exploited for the resource allocation algorithm design for SWIPT systems with multiuser scheduling and imperfect channel state information (CSI), respectively. Yet, the above works only consider single-antenna receivers, such that spatial multiplexing gains cannot be exploited, even if the transmitter is equipped with multiple antennas. Besides, the optimal resource allocation algorithm design for WPCNs has not been studied for practical non-linear EH models, yet.

Another challenging fundamental problem in multiuser WPCNs is how to achieve fairness in resource allocation. More specifically, wireless devices that are far away from the wireless power station can harvest considerably less energy compared to wireless devices in the proximity of the station \cite{JR:opp_and_ch_WPC}. Thus, distant wireless devices achieve a significantly smaller throughput, when they send their data to an information receiving station in the uplink using the wireless energy harvested in the downlink. In \cite{JR:WPC_Rui_Zhang}, system throughput maximization was considered in a multiuser WPCN, where the power station and the information receiving station were co-located. The system model proposed in \cite{JR:WPC_Rui_Zhang} gives rise to the ``doubly near-far" problem. The authors in \cite{JR:WPC_Rui_Zhang} tackled this problem by jointly optimizing the minimum user throughput in the system and the time allocation for the wireless powered users. In \cite{JR:sum_rate_mimo_WPCN_rank_one}, the authors extended the system model in \cite{JR:WPC_Rui_Zhang} to a multiuser multiple-input multiple-output (MIMO) WPCN. Thereby, the time allocation and the downlink and uplink precoding matrices were jointly optimized to maximize the uplink sum rate performance. However, the schemes in \cite{JR:WPC_Rui_Zhang, JR:sum_rate_mimo_WPCN_rank_one} may lead to a performance degradation in practical WPCNs, since their design was based on the linear EH model.  Moreover, perfect CSI knowledge was assumed in \cite{JR:WPC_Rui_Zhang, JR:sum_rate_mimo_WPCN_rank_one}, which may be too optimistic. Considering the fact that it is difficult to obtain perfect CSI in practice \cite{JR:Kwan_secure_imperfect,JR:Robust_error_models1}, due to CSI estimation and quantization errors \cite{JR:robust_secure_AN_cognitive_SWIPT}, it is important to take CSI uncertainties for the design of resource allocation algorithms into account. Imperfect CSI estimation in SWIPT systems has been considered in \cite{JR:Kwan_secure_imperfect}, \cite{JR:robust_secure_AN_cognitive_SWIPT}, \cite{JR:Robust_secure_miso_SWIPT_Feng} in different contexts. The authors of \cite{JR:Kwan_secure_imperfect} developed a robust beamforming algorithm for the minimization of the total transmit power of a secure multiuser SWIPT system. A similar robust resource allocation algorithm was proposed in \cite{JR:Robust_secure_miso_SWIPT_Feng}, with the objective of maximizing the secrecy rate of a SWIPT system. Both \cite{JR:Kwan_secure_imperfect} and \cite{JR:Robust_secure_miso_SWIPT_Feng} employ a deterministic model \cite{JR:Robust_error_models1,JR:Robust_error_models2,JR:CSI-determinisitic-model} for modeling the CSI uncertainty. In \cite{JR:robust_secure_AN_cognitive_SWIPT}, a robust beamforming algorithm was developed for secure multiple-input single-output (MISO) cognitive radio systems employing SWIPT, where the authors used both the deterministic and a probabilistic model to capture the impact of imperfect CSI knowledge. Additionally, the authors in \cite{JR:Robust_WPCN} proposed a joint design for robust beamforming and time allocation in a MISO WPCN with imperfect CSI knowledge at the power station. However, \cite{JR:Kwan_secure_imperfect}, \cite{JR:robust_secure_AN_cognitive_SWIPT}, \cite{JR:Robust_secure_miso_SWIPT_Feng}, \cite{JR:Robust_WPCN} assumed the conventional linear EH model for the end-to-end WET in the considered system architectures, which may lead to resource allocation mismatches and  performance degradation. Therefore, the design of robust resource allocation algorithms for SWIPT and WPCN systems that take into account both the CSI uncertainty and the non-linear EH characteristics is of high interest.

In this paper, we address the above issues. To this end, we consider a MIMO-WPCN, where multiple users harvest wireless energy from a dedicated power station and then send their information signals to a separate information receiving station. The main contributions of this paper are stated in the following: 
\begin{itemize}
\item[$\bullet$] We formulate the joint time allocation and power control algorithm design based on a non-linear EH model as a non-convex optimization problem. Two different system design objectives are considered, namely the maximization of the system sum throughput (max-sum) and the maximization of the minimum individual throughput (max-min) at each wireless powered user, respectively. Moreover, the proposed resource allocation algorithm designs take into account imperfect CSI knowledge and multiple-antenna transceivers. In order to solve the resulting difficult non-convex optimization problems, we apply several transformations. To this end, we first assume that the time duration $\tau_0$ of the WET period is fixed, and transform the original non-convex optimization problems into equivalent convex optimization problems. For a given $\tau_0$, the max-sum and max-min convex optimization problems are solved and an intermediate solution is obtained. Then, we perform a one-dimensional search across all possible WET time durations to obtain the value of $\tau_0$
that yields the maximum objective value and retrieve the corresponding optimal power control and time allocation for the wireless powered users. 
\item[$\bullet$] The proposed resource allocation algorithm designs take into account the non-linear characteristic of the end-to-end WET. In contrast, previous works in the literature that considered similar system architectures, e.g. \cite{JR:WPC_Rui_Zhang}, \cite{JR:sum_rate_mimo_WPCN_rank_one}, adopted the overly simplified linear EH model for the end-to-end WET. The linear EH model has been shown to be highly inaccurate with respect to practical EH circuits \cite{Letter_non_linear}, and using it for resource allocation design may lead to resource allocation mismatches and overall degradation of the system performance. 
\item[$\bullet$] We show that even with imperfect CSI knowledge, energy beamforming is optimal for WET in the considered MIMO-WPCN. Moreover, the optimal power allocation in the  WIT period has a water-filling structure. Besides, an analytical solution for the optimal time allocation is provided. 
 \item[$\bullet$] Computer simulations for both considered design objectives provide significant insights into the performance of MIMO-WPCNs employing the proposed resource allocation schemes. Specifically, the proposed resource allocation schemes achieve a significantly higher system performance compared to baseline resource allocation schemes designed based on the conventional linear EH model and are shown to be more robust against imperfect CSI knowledge than a non-robust benchmark scheme designed for perfect CSI. Furthermore, a comparison of the results obtained for the max-sum and the max-min schemes reveals a non-trivial trade-off between maximizing the system sum throughput and guaranteeing fairness in resource allocation among the wireless powered users in MIMO-WPCNs.\end{itemize}

The remainder of this paper is organized as follows. Section \ref{sect:System_model} introduces the system model and some preliminaries regarding the considered MIMO-WPCN, including the non-linear EH model and the CSI uncertainty model. In Section III, we formulate the max-sum and the max-min optimization problems. The solution of the optimization problems and the proposed resource allocation algorithms are presented in Section IV. In Section \ref{sect:simulation}, the performance of the proposed algorithms is evaluated via computer simulations. Conclusions are drawn in Section \ref{sect:conclusion}.

\textbf{Notation:}
$\mathbf{A}^H$, $\Tr(\mathbf{A})$, $\det(\mathbf{A})$, $\mathbf{A}^{-1}$, and $\Rank(\mathbf{A})$ represent the  Hermitian transpose, trace, determinant, inverse, and rank of  matrix $\mathbf{A}$, respectively; $\mathbf{A}\succeq \mathbf{0}$ indicates that $\mathbf{A}$ is a  positive semi-definite matrix; matrix $\mathbf{I}_{N}$
denotes the $N\times N$ identity matrix.  $\mathbb{C}^{N\times M}$ denotes the space of all $N\times M$ matrices with complex entries.
$\mathbb{H}^N$ represents the set of all $N$-by-$N$ complex Hermitian matrices.
$\norm{ \cdot }_2$, $\norm{ \cdot }_{\infty}$, and $\norm{\cdot}_{\mathrm{F}}$ denote the spectral norm, the infinity norm, and the Frobenius norm, respectively. The distribution of a circularly symmetric complex Gaussian (CSCG) vector with mean vector $\mathbf{x}$ and covariance matrix $\mathbf{\Sigma}$  is denoted by ${\cal CN}(\mathbf{x},\mathbf{\Sigma})$, and $\sim$ means ``distributed as".  $\cal E\{\cdot\}$ denotes statistical expectation and $\otimes$ stands for the Kronecker product. $\frac{\partial f({x})}{\partial x}$ represents the partial derivative of function $f(x)$ with respect to variable $x$. The gradient $\nabla_\mathbf{x} f(\mathbf{x})$ represents the partial derivative of function $f(\mathbf{x})$ with respect to the elements of vector $\mathbf{x}$. [$\mathbf{B}]_{a:b,c:d}$ returns a submatrix of $\mathbf{B}$ including the $a$-th to the $b$-th rows and the $c$-th to the $d$-th columns of $\mathbf{B}$; $\mathrm{vec} (\mathbf{B})$ results in a column vector, obtained by sequential stacking of the columns of matrix $\mathbf{B}$, and $\text{diag}(\mathbf{x})$ is a diagonal matrix with the elements of vector $\mathbf{x}$ on the main diagonal.
\section{System Model and Preliminaries} \label{sect:System_model}

In this section, we define the channel, energy harvesting, and CSI models, which are adopted for resource allocation algorithm design. 

\subsection{Channel Model}

\begin{figure}[t]
\centering
\includegraphics[width=4.2 in]{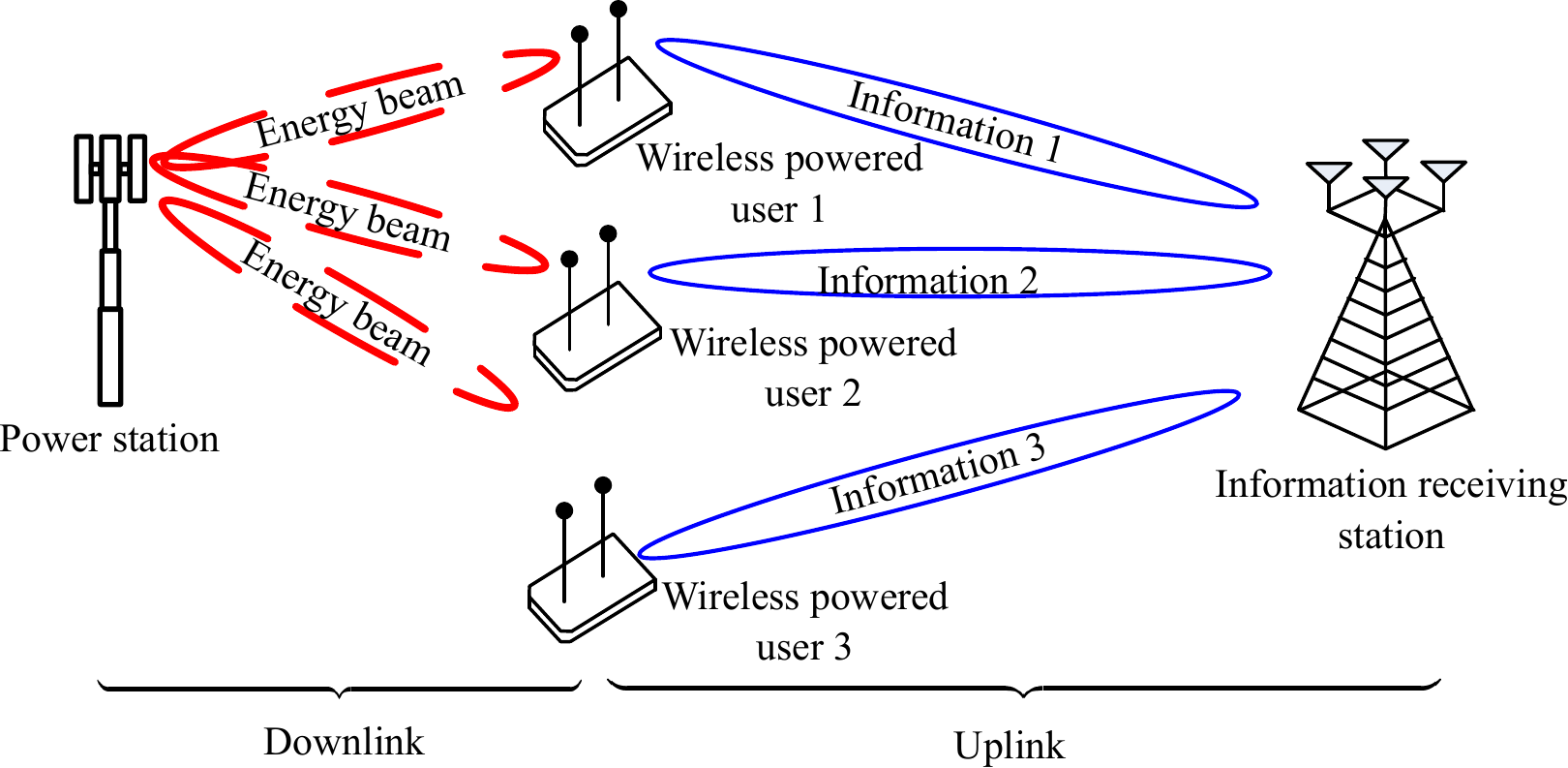}
\caption{A downlink wireless powered communication system with $K=3$ multiple-antenna users.}
\label{fig:system_model}
\end{figure}

\begin{figure}[t]\vspace*{-2mm}
\centering
\includegraphics[width=3.6 in]{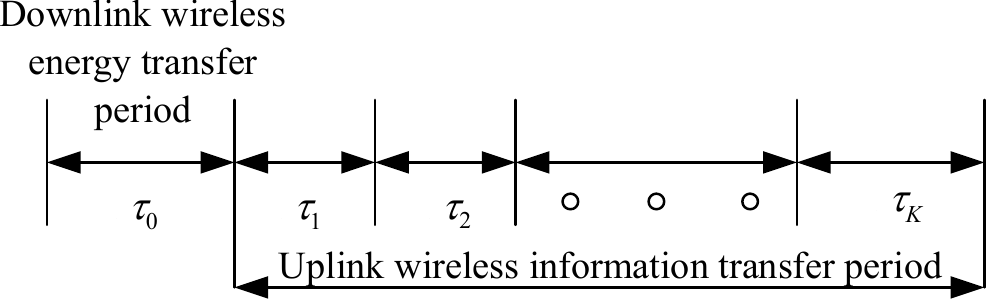}
\caption{Wireless energy and information transfer protocol.}
\label{fig:system_model2}
\end{figure}

In the considered MIMO-WPCN, a power station delivers wireless energy  to  $K$  wireless powered users in the downlink to facilitate the users' information transfer to an information receiving station in the uplink, cf. Figure \ref{fig:system_model}.
To avoid the ``doubly near-far'' problem \cite{JR:WPC_Rui_Zhang,CN:User_Coop_WPCN}, we assume different stations for WET and wireless information transfer (WIT) \cite{JR:WPC_QQ}.
 The power station, the wireless powered users, and the information receiving station are each equipped with $N_\mathrm{T}\geq1$, $N_{\mathrm{U}_k}\geq1$,$\forall k$, and $N_\mathrm{R}\geq1$ antennas, respectively. Moreover, for simplicity, we assume that all transceivers operate in the same frequency band using time division multiple access. In the considered network, we adopt the ``harvest-then-transmit" protocol \cite{JR:WPC_Rui_Zhang,JR:WPC_QQ}. Specifically, the transmission is divided into two periods, namely a downlink WET period and an uplink WIT period, cf. Figure \ref{fig:system_model2}. In the WET period, the power station\footnote{We note that having one multiple-antenna power station is mathematically equivalent to having multiple power stations that are connected and share the power resources.} sends a vector of energy signals to the $K$ wireless powered users for energy harvesting.
 Subsequently, the users utilize all of the energy harvested during the WET period to transmit their information signals to the information receiving station in the WIT period.  The time for WET and the transmission time of each wireless powered user during the WIT period can be optimized \cite{JR:WPC_Rui_Zhang}. We assume that each wireless powered user is equipped with a rechargeable battery which has a sufficiently large capacity to store the amount of energy harvested during the WET period \cite{JR:Kwan_secure_imperfect}, \cite{JR:WPC_Rui_Zhang}, \cite{JR:WPC_QQ}. Furthermore, we assume a frequency flat slowly time-varying fading channel in both downlink and uplink. The instantaneous received signal at wireless powered user $k\in\{1,\ldots,K\}$ is given by
\begin{eqnarray}
\mathbf{y}_{\mathrm{EH}_k}&=&\mathbf{G}_k^H\mathbf{v}+\mathbf{n}_{\mathrm{EH}_k},
\end{eqnarray} where  $\mathbf{v}\in\mathbb{C}^{N_{\mathrm{T}}\times1}$ is the random energy signal vector adopted in the downlink for WET, with covariance matrix $\mathbf{V} = \mathcal{E} \{ \mathbf{v}\mathbf{v}^H \}$. The channel matrix between the power station and wireless powered user $k$ is denoted by  $\mathbf{G}_k\in\mathbb{C}^{N_{\mathrm{T}}\times N_{\mathrm{U}_k}}$
 and captures the joint effect of pathloss and multipath fading. Vector $\mathbf{n}_{\mathrm{EH}_k}\sim{\cal CN}(\zero,\sigma_{\mathrm{s}_k}^2\mathbf{I}_{N_{\mathrm{U}_k}})$ represents the additive white Gaussian noise (AWGN) at
wireless powered user $k$ where $\sigma_{\mathrm{s}_k}^2$ denotes the noise variance at each antenna of the user. Then, in the uplink WIT period, all $K$ wireless powered users exploit the energy harvested during the WET period to transmit independent information signals to the information receiving station. Thereby, wireless powered user $k$ is allocated $\tau_k$
   amount of time for uplink transmission. The signal received from wireless powered user $k$ at the information receiving station is given by
\begin{equation}
\mathbf{y}_{\mathrm{IR}_k}=\mathbf{H}_k^H\mathbf{Q}_k \mathbf{s}_k+ \mathbf{n},\,\,  \forall k\in\{1,\dots,K\},
\end{equation}
where $\mathbf{H}_k\in\mathbb{C}^{N_{\mathrm{U}_k}\times N_{\mathrm{R}}}$  is the channel matrix between wireless powered user $k$ and the information receiving station. Vector $\mathbf{s}_k\in\mathbb{C}^{N_{\mathrm{s}_k}\times 1}$ comprising $N_{\mathrm{s}_k}$ information-carrying symbols is the information signal vector of wireless powered user $k$, and  $\mathbf{Q}_k\in\mathbb{C}^{N_{\mathrm{U}_k}\times N_{\mathrm{s}_k}}$ is the precoding matrix adopted at wireless powered user $k$ for WIT. 
$\mathbf{n}\sim{\cal CN}(\zero,\sigma_{n}^2\mathbf{I}_{N_{\mathrm{R}}})$ is the AWGN vector at the information receiving station and $\sigma_{n}^2$ denotes the corresponding noise variance. Without loss of generality, we assume that ${\cal E}\{\mathbf{s}_{k}\mathbf{s}_{k}^H\}=\mathbf{I}_{N_{\mathrm{s}_k}},\forall k\in\{1,\ldots,K\}$, where $N_{\mathrm{s}_k}\leq \min\{N_{\mathrm{U}_k},N_{\mathrm{R}}\}$.

\subsection{Energy Harvesting Model}
In the literature,  the total energy harvested by wireless powered user $k$ during the wireless charging phase is typically modeled by the following linear model
 \cite{CN:WIPT_fundamental}--\cite{JR:OFDM_relay_SWIPT}:
\begin{eqnarray}\label{eqn:linear_model}
\Phi_{\mathrm{EH}_k}^{\mathrm{Linear}}=\eta_k P_{\mathrm{EH}_k},\quad
P_{\mathrm{EH}_k}=\Tr\Big(\mathbf{G}_k^H\mathbf{V}\mathbf{G}_k\Big),
\end{eqnarray}
where $P_{\mathrm{EH}_k}$ is the total received RF power at wireless powered user $k$, and $0\leq\eta_k\leq1$ is the constant energy conversion efficiency for converting RF energy to electrical energy at wireless powered user $k$. We emphasize that in this linear EH model, the energy conversion efficiency is independent of the input power level at the wireless powered user. In other words, the total harvested energy at the energy harvesting receiver is linearly proportional to the received RF power. However, as was shown in various experiments, practical RF-based EH circuits have a non-linear end-to-end WET characteristic  \cite{JR:Energy_harvesting_circuit}\nocite{JR:EH_measurement_1}--\cite{CN:EH_measurement_2}. In particular, the RF energy conversion efficiency first improves as the input power increases, but for high input powers there is a diminishing return and a limitation on the maximum harvested energy.
\begin{figure}[t]
\centering\vspace*{-8mm}
\includegraphics[width=3.5 in]{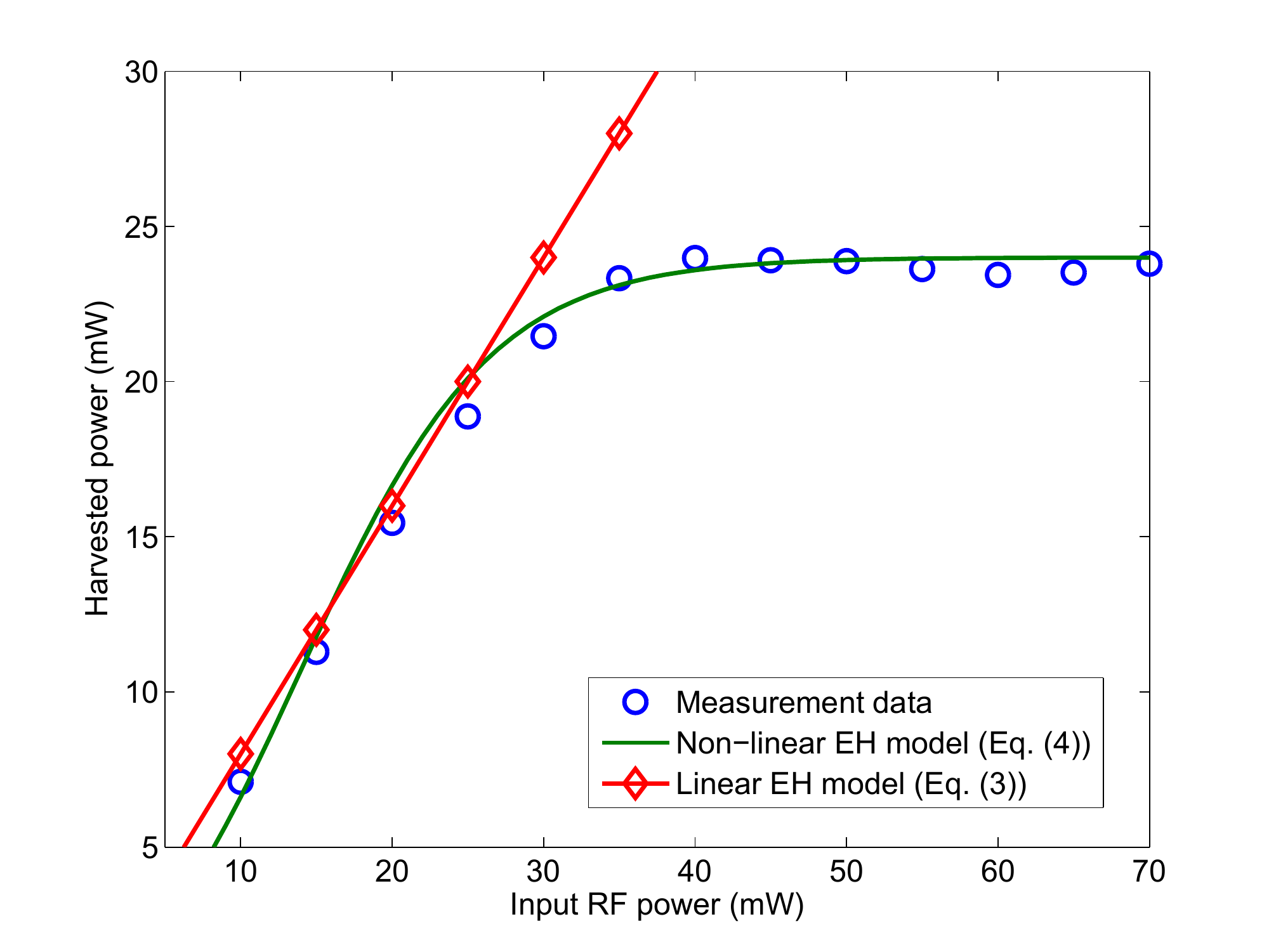}
\caption{A comparison between the harvested power according to the non-linear EH model in \eqref{eqn:EH_non_linear}, the linear EH model in \eqref{eqn:linear_model}, and measurement data provided for a practical EH circuit in \cite{CN:EH_measurement_2}.  The parameters $a_k = 150$, $b_k = 0.014$, and $M_k = 0.024$ in \eqref{eqn:EH_non_linear} were obtained using a standard curve fitting tool.}
\label{fig:comparsion_measurment}\vspace*{-4mm}
\end{figure}
Thus, employing the linear EH model to characterize the end-to-end WET for resource allocation algorithm design may lead to a suboptimal performance.
As was recently shown for SWIPT systems in \cite{ICC_non_linear} and \cite{SPAWC_non_linear}, a non-linear EH model reflecting the non-linearity of practical EH circuits can avoid the resource allocation mismatches arising for the traditional linear EH model. However, the impact of adopting a practical non-linear EH model for resource allocation algorithm design for WPCNs has not been studied, yet. Here, we adopt the non-linear EH model from \cite{Letter_non_linear} and employ it for characterizing the RF-to-DC power transfer at the wireless powered user terminals in the WET phase. The non-linear EH model from \cite{Letter_non_linear} is given by:
 \begin{eqnarray}\label{eqn:EH_non_linear}
 \hspace*{-5mm}\Phi_{\mathrm{EH}_k}^{\mathrm{Practical}}\hspace*{-2mm}&=&\hspace*{-2mm}
 \frac{[\Psi_{\mathrm{EH}_k}^{\mathrm{Practical}}
 \hspace*{-0.5mm}- \hspace*{-0.5mm}M_k\Omega_k]}{1-\Omega_k},\, \Omega_k=\frac{1}{1+\exp(a_kb_k)},\notag\\
 \hspace*{-5mm}\Psi_{\mathrm{EH}_k}^{\mathrm{Practical}}\hspace*{-2mm}&=&\hspace*{-2mm} \frac{M_k}{1+\exp\Big(\hspace*{-0.5mm}-a_k(P_{\mathrm{EH}_k}-\hspace*{-0.5mm}b_k)\Big)}.
  \end{eqnarray}
Here, $\Phi_{\mathrm{EH}_k}^{\mathrm{Practical}}$ is the total harvested energy at wireless powered user $k$ and $\Psi_{\mathrm{EH}_k}^{\mathrm{Practical}}$ is the conventional logistic function with respect to the received RF power $P_{\mathrm{EH}_k}$.
%
By adjusting the parameters $M_k$, $a_k$, and $b_k$, the non-linear EH model is able to capture the joint effects of various non-linear phenomena caused by hardware limitations \cite{Letter_non_linear}. In particular, $M_k$ denotes the maximum power that the EH receiver can harvest, as the EH circuit saturates if the received RF power is exceedingly large, while $a_k$ and $b_k$ can account for physical hardware phenomena, such as circuit sensitivity limitations and leakage currents \cite{JR:Energy_harvesting_circuit}\nocite{JR:EH_measurement_1}--\cite{CN:EH_measurement_2}.

Figure \ref{fig:comparsion_measurment} illustrates that the proposed non-linear EH model in \eqref{eqn:EH_non_linear} closely matches with experimental results reported in \cite{CN:EH_measurement_2} for the wireless power harvested by a practical EH circuit. Besides, Figure \ref{fig:comparsion_measurment} also reveals the limitations of the conventional linear EH model in \eqref{eqn:linear_model} in accurately modeling non-linear EH circuits.
\subsection{Channel State Information}
 In this paper, we assume that there is a central unit (e.g. the power station or the information receiving station) collecting the CSI of all the wireless links for computation of the resource allocation policy. Since the considered wireless channels change  slowly over time, the CSI of all the links becomes outdated at the central unit  during transmission. In the literature, there are two different approaches to capture the impact of imperfect CSI knowledge, which differ in the way the CSI errors are modeled. The first approach is based on a deterministic model, while the second approach is based on a probabilistic model, where the CSI errors are modeled by continuous random variables following a certain distribution \cite{JR:secrecy_capacity_prob_det_models}. We note that there is no restriction on the maximum error magnitude in the probabilistic model. In fact, the probabilistic model can be converted to the deterministic model under some general conditions \cite[Proposition 1]{JR:secrecy_capacity_prob_det_models}. As a result, in this paper, we adopt the deterministic model \cite{JR:Robust_error_models1,JR:Robust_error_models2,JR:CSI-determinisitic-model}
 in order to capture the impact of imperfect CSI knowledge and to isolate the specific
channel estimation method used from the resource allocation algorithm
design. According to this model, the  CSI between the power station and wireless powered user $k$ and between wireless powered user $k$ and the information receiving station  can be modeled as
\begin{eqnarray}\label{eqn:outdated_CSI}
\mathbf{G}_k&=&\mathbf{\widehat G}_k + \Delta\mathbf{G}_k,\,   \forall k\in\{1,\ldots,K\},\label{eqn:outdated_CSI-set-1a} \\
{\bm\Xi }_k&\triangleq& \Big\{\Delta\mathbf{G}_k\in \mathbb{C}^{N_{\mathrm{T}}\times N_{\mathrm{U}_k}}  :\norm{\Delta\mathbf{G}_k}_\mathrm{F}  \le \upsilon_k\Big\},\forall k,\label{eqn:outdated_CSI-set-1b}\,\mbox{   and}\\
\mathbf{H}_k&=&\mathbf{\widehat H}_k + \Delta\mathbf{H}_k,\,   \forall k\in\{1,\ldots,K\}, \label{eqn:outdated_CSI-set-2a}\\
{\bm\Lambda }_k&\triangleq& \Big\{\Delta\mathbf{H}_k\in \mathbb{C}^{N_{\mathrm{U}_k}\times {N_{\mathrm{R}}}}  :\norm{\Delta\mathbf{H}_k}_\mathrm{F}  \le \rho_k\Big\}, \forall k, \label{eqn:outdated_CSI-set-2b}
\end{eqnarray}
respectively, where $\mathbf{\widehat G}_k$ and $\mathbf{\widehat H}_k$ are the  estimates of channel matrices $\mathbf{ G}_k$ and $\mathbf{H}_k$, respectively,  at the central unit. Matrices $\Delta\mathbf{G}_k$ and $\Delta\mathbf{H}_k$  represent the channel uncertainty and capture the joint effects of channel estimation errors and the time varying nature of the associated channels. In particular,  the continuous sets ${\bm\Xi}_k$ and ${\bm\Lambda }_k$ in (\ref{eqn:outdated_CSI-set-1b}) and \eqref{eqn:outdated_CSI-set-2b}, respectively,  define the continuous spaces spanned by all possible channel uncertainties. Constants $\upsilon_k$ and $\rho_k$ denote the maximum value of the norm of the CSI estimation error matrices  $ \Delta\mathbf{G}_k$ and $ \Delta\mathbf{H}_k$ for wireless powered user $k$. In practice, the values of $\upsilon_k$ and $\rho_k$ depend on the coherence time of the associated channels, the duration of the scheduling slot, and the specific channel estimation schemes. We note that the duration of a scheduling slot is typically much longer than the duration of an information packet. The adopted CSI model\footnote{We note that the general model adopted for the CSI estimation errors in the considered MIMO-WPCN allows us to isolate the resource allocation design from specific implementation parameters such as the duplexing method.} in \eqref{eqn:outdated_CSI-set-1a}-\eqref{eqn:outdated_CSI-set-2b} takes into account the imperfect CSI at the central unit for performing resource allocation. On the other hand, we assume that  pilot sequences are embedded in the information packets such that the information receiving station is able to frequently update and refine the
CSI estimates during information transmission \cite{JR:Kwan_secure_imperfect}, \cite{JR:Robust_error_models1}. Thus, we assume that perfect CSI at the receiver (CSIR) is available for coherent information decoding at the information receiving station.

\section{Resource Allocation Problem Formulation}

In the following, we formulate the optimization problem for maximization of the total system throughput, i.e., the max-sum resource allocation problem, and the optimization problem for maximization of the minimum individual throughput at each wireless powered user, i.e., the max-min resource allocation problem. 

\subsection{Max-sum Problem Formulation}\label{sect:problem_formulation_solution}
In this section, we present the problem formulation for the max-sum resource allocation algorithm design. The goal of the resource allocation is to jointly optimize the time allocation and power control for maximization of the sum throughput at the information receiving station for the considered non-linear EH model. The resource allocation policy, $\{{\bm \tau}, \mathbf{V}, \mathbf{Q}_{k}\}$, for maximizing the total system throughput, can be obtained by solving
\begin{eqnarray}\label{eqn:TP_maximization}
\underset{ \mathbf{V}\in \mathbb{H}^{N_{\text{\tiny{T}}}}, \mathbf{Q}_{k}\in \mathbb{H}^{N_{\text{\tiny U}_k}},{\tau_k}}{\maxo}\,\, \hspace*{-2mm}&& \sum_{k=1}^K  \min_{\Delta \mathbf{H}_k\in {\bm \Lambda}_k}\, \tau_k\log_2\,\det\Big(\mathbf{I}_{N_{\mathrm{R}}}+\frac{1}{\sigma_n^2}\mathbf{H}_k^H\mathbf{Q}_k\mathbf{H}_k\Big)\\
\hspace*{-2mm}\mathrm{subject\,\,to}\,\, &&\mathrm{C1}:\,\,\Tr(\mathbf{V})\le P_{\max},\notag\\
\hspace*{-2mm}&&\mathrm{C2}:\,\, \tau_0 +\sum_{k=1}^K \tau_k \leq T_{\max},\notag\\
\hspace*{-2mm}&&\mathrm{C3}:\,\, T_{\max} P_{\mathrm{c}_k}+\tau_k \Tr(\mathbf{Q}_{k})\varepsilon_k \leq \min_{\Delta \mathbf{G}_k\in {\bm \Xi}_k} \tau_0\Phi_{\mathrm{EH}_k}^{\mathrm{Practical}},\,\forall k,\notag\\
\hspace*{-2mm}&&\mathrm{C4}:\,\, \tau_r \ge 0, \forall r\in\{0,1,\ldots,K\},\notag\\
\hspace*{-2mm}&&\mathrm{C5}:\,\, \mathbf{V}\succeq \zero,\notag\\
\hspace*{-2mm}&&\mathrm{C6}:\,\, \mathbf{Q}_k\succeq \zero,\forall k\in\{1,\ldots,K\}.\notag
\end{eqnarray}
Here, $\bm{\tau} = \{ \tau_0, \tau_1,\ldots, \tau_K \} $ is the time allocation vector, comprising both the downlink WET time $\tau_0$ and the corresponding uplink WIT periods $\tau_k$, $\forall k$.
$\mathbf{Q}_{k}$ is the covariance matrix of the information signal of wireless powered user $k$. By exploiting the channel model in \eqref{eqn:outdated_CSI-set-2a}, the objective function in \eqref{eqn:TP_maximization} takes into account the CSI
 uncertainty set ${\bm \Lambda}_k$ to provide robustness against CSI imperfection. Constants $P_{\max}$ and $T_{\max}$  in constraints C1 and C2 represent the maximum transmit power of the power station and the maximum duration of a transmission slot, respectively. Moreover, constraint C3 is imposed such that, for a given CSI uncertainty set ${\bm \Xi}_k$,  the maximum available energy for wireless powered user $k$ for uplink WIT is limited by the harvested energy during the downlink WET period $\tau_0$ in the corresponding time slot. The amount of harvested power at wireless powered user $k$ is computed based on the practical non-linear EH model in \eqref{eqn:EH_non_linear}. The minimization on the right-hand side of constraint C3 is performed with respect to all possible CSI estimation errors $\Delta \mathbf{G}_k$ of the CSI uncertainty set $\mathbf{\Xi}_k$ for the estimation of the channel between the power station and wireless powered user $k$, i.e., $\mathbf{G}_k$. Hence, constraint C3 ensures that the optimum performance in the considered MIMO-WPCN is guaranteed even for the worst-case CSI estimation error according to CSI uncertainty set ${\bm \Xi}_k$. $P_{\mathrm{c}_k}$ in constraint C3 is the constant circuit power consumption. Besides, to capture
the power inefficiency of the power amplifiers, we introduce in C3 a linear multiplicative constant $\varepsilon_k>1$ for the power radiated by wireless powered user $k$. For example, if $\varepsilon_k=5 $, then for every $1$ Watt of power radiated in the
RF, wireless powered user $k$ consumes $5$ Watt of power which leads to a power
amplifier efficiency of $20\%$.  C4 is the non-negativity constraint for the transmission period $\tau_k$ of wireless powered user $k$, $\forall k$. Constraints C5, C6, $\mathbf{V}\in \mathbb{H}^{N_\mathrm{T}} $, and $\mathbf{Q}_k\in \mathbb{H}^{N_{\mathrm{U}_k}} $ constrain matrices $\mathbf{V}$ and $\mathbf{Q}_k$ to be positive semi-definite Hermitian matrices.

\subsection{Max-min Problem Formulation}\label{sect:problem_formulation_max_min}

Resource allocation algorithms focusing solely on maximizing the sum throughput usually result in an unfair resource allocation, since users with good channel conditions consume most of the system resources \cite{JR:gen_opt_framework_WPNC} which leads to the starvation of users with poor channel conditions. Motivated by this fact, we also formulate a fair resource allocation optimization problem for the considered MIMO-WPCN that aims to maximize the minimum throughput across the wireless powered users in the system. 
The resource allocation policy for the max-min fairness optimization problem, $\{{\bm \tau}, \mathbf{V}, \mathbf{Q}_{k}\}$, can be obtained by solving the following optimization problem:
\begin{eqnarray}\label{eqn:TP_max_min}
\underset{ \mathbf{V}\in \mathbb{H}^{N_{\text{\tiny{T}}}}, \mathbf{Q}_{k}\in \mathbb{H}^{N_{\text{\tiny U}_k}},{\tau_k}}{\maxo}\,\, \hspace*{-2mm}&& \underset{k}{\min}  \Bigg(  \min_{\Delta \mathbf{H}_k\in {\bm \Lambda}_k}\, \tau_k\log_2\,\det\Big(\mathbf{I}_{N_{\mathrm{R}}}+\frac{1}{\sigma_n^2}\mathbf{H}_k^H\mathbf{Q}_k\mathbf{H}_k\Big)\Bigg)\\
\hspace*{-2mm}\mathrm{subject\,\,to}\,\, &&\mathrm{C1}:\,\,\Tr(\mathbf{V})\le P_{\max},\notag\\
\hspace*{-2mm}&&\mathrm{C2}:\,\, \tau_0 +\sum_{k=1}^K \tau_k \leq T_{\max},\notag\\
\hspace*{-2mm}&&\mathrm{C3}:\,\, T_{\max} P_{\mathrm{c}_k}+\tau_k \Tr(\mathbf{Q}_{k})\varepsilon_k \leq \min_{\Delta \mathbf{G}_k\in {\bm \Xi}_k} \tau_0\Phi_{\mathrm{EH}_k}^{\mathrm{Practical}},\,\forall k,\notag\\
\hspace*{-2mm}&&\mathrm{C4}:\,\, \tau_r \ge 0, \forall r\in\{0,1,\ldots,K\},\notag\\
\hspace*{-2mm}&&\mathrm{C5}:\,\, \mathbf{V}\succeq \zero,\notag\\
\hspace*{-2mm}&&\mathrm{C6}:\,\, \mathbf{Q}_k\succeq \zero,\forall k\in\{0,1,\ldots,K\}.\notag
\end{eqnarray}

The objective function of the optimization problem in \eqref{eqn:TP_max_min} maximizes the minimum individual throughput among all wireless powered users, while taking into account the imperfect CSI knowledge. The optimization problem in \eqref{eqn:TP_max_min} ensures fairness among the different wireless powered users in the sense that each of them will achieve at least the minimum individual throughput. Besides, the constraint set of problem \eqref{eqn:TP_max_min} is identical to the constraint set of the sum throughput optimization problem in \eqref{eqn:TP_maximization}.

\section{Solution of the Optimization Problems}

The optimization problems in \eqref{eqn:TP_maximization} and \eqref{eqn:TP_max_min} are non-convex optimization problems that involve infinitely many constraints. The non-convexity of the problems arises from constraint C3 and the objective function. Specifically, constraint C3 couples the optimization variables $\tau_k$ and $\mathbf{Q}_k$, and its right-hand side is a quasi-concave function. If the resource allocation was based on the linear EH model in \eqref{eqn:linear_model}, the right-hand side of constraint C3 would be affine. Thereby, we would not face the difficulties in solving the optimization problems that rise from the fractional nature of the non-linear EH model. Another difficulty is the continuity of the channel uncertainty set, which introduces an infinite number of constraints in C3. Similarly, the channel uncertainty  introduces an infinite number of possibilities for the objective functions. In addition, even if perfect CSI was available, the objective functions in their original formulations would not be jointly concave with respect to optimization variables $\tau_k$ and $\mathbf{Q}_k$.  To obtain a tractable problem formulation and to solve the problems by using efficient convex optimization tools, we introduce several transformations for problems \eqref{eqn:TP_maximization} and \eqref{eqn:TP_max_min} in the following. Specifically, we will first present the detailed solution steps for problem \eqref{eqn:TP_maximization}. Subsequently, we will extend the concept to efficiently solve problem \eqref{eqn:TP_max_min}.

\subsection{Transformation of Constraint C3}

In order to handle the quasi-concavity of constraint C3, we solve the optimization problem in \eqref{eqn:TP_maximization} for a fixed constant $\tau_0$ and obtain the corresponding resource allocation policy. Then, using a one-dimensional search, we find the optimal value of the optimization problem and the corresponding $\tau_0$ for that instant.
Furthermore, to handle the infinitely many constraints due to the CSI error uncertainty set, we first introduce an auxiliary optimization variable $\theta_k$, and rewrite constraint C3 in \eqref{eqn:TP_maximization} in the following equivalent form:
\begin{eqnarray}
\hspace*{-2mm}&&\hspace*{-10mm}\mathrm{C3a}:\,\, T_{\max} P_{\mathrm{c}_k}+\tau_k\Tr(\mathbf{Q}_{k})\varepsilon_k\leq \tau_0
 \frac{\frac{M_k}{1+\exp\big(\hspace*{-0.5mm}-a_k(\theta_k-\hspace*{-0.5mm}b_k)\big)}
 \hspace*{-0.5mm}- \hspace*{-0.5mm}M_k\Omega_k}{1-\Omega_k},\,\forall k,\\
 \hspace*{-2mm}&&\hspace*{-10mm}\mathrm{C3b}:\,\, \theta_k\leq \min_{ \norm{\Delta \mathbf{G}_k}_F^2 \leq \upsilon_k^2} \Tr\Big(\mathbf{G}_k^H\mathbf{V}\mathbf{G}_k\Big),\,\forall k.
\end{eqnarray}
It can be shown that for the optimal solution, constraint C3b is satisfied with equality.
To facilitate the derivation of the solution, we further transform constraint C3b into a linear matrix inequality (LMI) using the following theorem:
\begin{Lem}[S-Procedure \cite{book:convex}] Let a function $f_m(\mathbf{x}),m\in\{1,2\},$ be defined as
\begin{eqnarray}
f_m(\mathbf{x})=\mathbf{x}^H\mathbf{A}_m\mathbf{x}+2 \mathrm{Re} \{\mathbf{b}_m^H\mathbf{x}\}+c_m,
\end{eqnarray}
where $\mathbf{A}_m\in\mathbb{H}^N$, $\mathbf{b}_m\in\mathbb{C}^{N\times 1}$, and $c_m\in\mathbb{R}$. Then, the implication $f_1(\mathbf{x})\le 0\Rightarrow f_2(\mathbf{x})\le 0$  holds if and only if there exists an $\omega\ge 0$ such that
\begin{eqnarray}\omega
\begin{bmatrix}
       \mathbf{A}_1 & \mathbf{b}_1          \\
       \mathbf{b}_1^H & c_1           \\
           \end{bmatrix} -\begin{bmatrix}
       \mathbf{A}_2 & \mathbf{b}_2          \\
       \mathbf{b}_2^H & c_2           \\
           \end{bmatrix}          \succeq \mathbf{0},
\end{eqnarray}
provided that there exists a point $\mathbf{\hat{x}}$ such that $f_k(\mathbf{\hat{x}})<0$.
\end{Lem}

To apply Lemma 1 to constraint C3b, we rewrite \eqref{eqn:outdated_CSI-set-1b} and reformulate constraint C3b.  In particular,
\begin{equation}
\Delta \mathbf{g}_k^H \Delta \mathbf{g}_k \leq \upsilon_k^2 \Longrightarrow \mathbf{\widehat g}_k^H \bm{\mathcal{V}} \mathbf{\widehat g}_k + 2\mathrm{Re} \{ \mathbf{\widehat g}_k^H \bm{\mathcal{V}}  \Delta \mathbf{g}_k \} + \Delta \mathbf{g}_k^H \bm{\mathcal{V}} \Delta \mathbf{g}_k - \theta_k \geq 0,
\end{equation}
holds if and only if there exist $\omega_k, \forall k$, such that the following LMI constraints hold:
\begin{eqnarray}\mathbf{\Upsilon} (\bm{\mathcal{V}}, \omega_k, \theta_k) && \triangleq
             \begin{bmatrix}
       \omega_k \mathbf{I}_{N_\mathrm{T}  N_{\mathrm{U}_k} } & \mathbf{0}\\
       \mathbf{0} & -\omega_k \upsilon_k^2 -\theta_k\\
           \end{bmatrix} + \mathbf{U}_{\widehat{\mathbf{g}}_k}^H\bm{\mathcal{V}}\mathbf{U}_{\widehat{\mathbf{g}}_k} \succeq \mathbf{0}, \, \forall k,
\end{eqnarray}
where $\mathbf{\widehat g}_k = \mathrm{vec}(\mathbf{\widehat G}_k), \Delta \mathbf{g}_k = \mathrm{vec}(\Delta \mathbf{G}_k)$,  $\bm{\mathcal{V}} = \mathbf{I}_{N_{\mathrm{U}_k}} \otimes \mathbf{V}$, and $ \mathbf{U}_{\widehat{\mathbf{g}}_k} = [\mathbf{I}_{N_\mathrm{T}  N_{\mathrm{U}_k} } \,\, \mathbf{\widehat g}_k], \forall k$.
Finally, we introduce an auxiliary optimization variable $\mathbf{\widetilde Q}_k$= $\mathbf{Q}_k\tau_k, \forall k$, to decouple the optimization variables in constraint C3a. Then, the reformulated optimization problem \eqref{eqn:TP_maximization} is given by:
               \begin{eqnarray}\label{eqn:TP_maximization_16}
\underset{ \mathbf{V}\in \mathbb{H}^{N_{\text{\tiny{T}}}},
{\mathbf{\widetilde Q}_k}\in \mathbb{H}^{N_{\text{\tiny U}_k}},{\tau_k}, \theta_k}{\maxo}\,\, \hspace*{-2mm}&& \sum_{k=1}^K  {\min_{\Delta \mathbf{H}_k\in {\bm \Lambda}_k}}\, \tau_k\log_2\,\det\Big(\mathbf{I}_{N_{\mathrm{R}}}+\frac{1}{\sigma_n^2}\mathbf{H}_k^H
{\frac{\mathbf{\widetilde Q}_k}{\tau_k}}\mathbf{H}_k\Big)\\
\hspace*{-2mm}\mathrm{subject\,\,to}\,\, &&\mathrm{C1}, \mathrm{C2}, \notag\\
\hspace*{-2mm}&&\mathrm{C3a}:\,\, T_{\max} P_{\mathrm{c}_k}+\Tr(\mathbf{\widetilde Q}_{k})\varepsilon_k\leq \tau_0 \frac{\frac{M_k}{1+\exp\big(\hspace*{-0.5mm}-a_k(\theta_k-\hspace*{-0.5mm}b_k)\big)}
 \hspace*{-0.5mm}- \hspace*{-0.5mm}M_k\Omega_k}{1-\Omega_k} ,\,\forall k,\notag\\
\hspace*{-2mm}&&\mathrm{C3b}:\,\, \mathbf{\Upsilon} (\bm{\mathcal{V}}, \omega_k, \theta_k) \succeq \mathbf{0} ,\,\forall k,\notag\\
\hspace*{-2mm}&&\mathrm{C4}, \mathrm{C5}, \notag \\
\hspace*{-2mm}&&\mathrm{C6}:\,\, \mathbf{\widetilde Q}_k\succeq \zero,\forall k, \,\, \mathrm{C7}:\,\, \omega_k \geq 0,\,\forall k.\notag 
\end{eqnarray}
The objective function in \eqref{eqn:TP_maximization_16} is jointly concave with respect to optimization variables $\mathbf{\widetilde Q}_{k}$ and $\tau_k$. Besides, constraint C3a is an affine function with respect to $\mathbf{\widetilde Q}_{k}$ for a given $\tau_0$ which yields a convex constraint set for problem \eqref{eqn:TP_maximization}. 

\subsection{Transformation of the Objective Function}
The remaining difficulties in solving the optimization problem in \eqref{eqn:TP_maximization} efficiently arise from the objective function. In order to tackle this challenge, we transform the objective function in the following. Due to the employed model for the CSI uncertainty set, i.e., the Frobenius norm $\norm{\cdot}_\mathrm{F}$, the objective function is intractable in its current form. On the other hand, the results in \cite{JR:Robust_MIMO_uncertainty} can be useful to transform the problem when the CSI uncertainty is modeled with respect to the spectral norm, i.e., the $\norm{\cdot}_2$-norm. In this context, we invoke the following inequality \cite{JR:Robust_MIMO_uncertainty,JR:MIMO:uncertainty_norm_depend}:
       \begin{eqnarray}\label{eqn:norm_inequality}
        \norm{\Delta \mathbf{H}_k}_2  \leq \norm{\Delta \mathbf{H}_k}_{\mathrm{F}} \leq \sqrt{\min(N_{\mathrm{U}_k},N_{\mathrm{R}})} \, \norm{\Delta \mathbf{H}_k}_2.
       \end{eqnarray}
 In the following, in order to design a computationally efficient resource allocation algorithm, we focus on a lower bound of the objective function:
     \begin{eqnarray}
     \underset{ \mathbf{V}\in \mathbb{H}^{N_{\text{\tiny{T}}}},
{\mathbf{\widetilde Q}_k}\in \mathbb{H}^{N_{\tiny \mathrm{U}_k}},{\tau_k}, \theta_k}{\maxo}\,\, \hspace*{-2mm}&& \sum_{k=1}^K  {\min_{\norm{\Delta \mathbf{H}_k}_{\mathrm{F}} \le \rho_k}}\, \tau_k\log_2\,\det\Big(\mathbf{I}_{\mathrm{R}}+\frac{1}{\sigma_n^2}\mathbf{H}_k^H
{\frac{\mathbf{\widetilde Q}_k}{\tau_k}}\mathbf{H}_k\Big)\\
\stackrel{\mathrm{(a)}}{\ge} \underset{ \mathbf{V}\in \mathbb{H}^{N_{\text{\tiny{T}}}},
{\mathbf{\widetilde Q}_k}\in \mathbb{H}^{N_{\tiny \mathrm{U}_k}},{\tau_k}, \theta_k}{\maxo}\,\, \hspace*{-2mm}&& \sum_{k=1}^K  {\min_{\norm{\Delta \mathbf{H}_k}_2\le \rho_k}}\, \tau_k\log_2\,\det\Big(\mathbf{I}_{\mathrm{R}}+\frac{1}{\sigma_n^2}\mathbf{H}_k^H
{\frac{\mathbf{\widetilde Q}_k}{\tau_k}}\mathbf{H}_k\Big),
 \label{eqn:lower_bound}
\end{eqnarray}
where $\mathrm{(a)}$  is due to \eqref{eqn:norm_inequality}. Then, with the lower bound of the objective function in \eqref{eqn:lower_bound}, the optimization problem in \eqref{eqn:TP_maximization_16} can be transformed to:
 \begin{eqnarray}
 \underset{ \substack{ \mathbf{V}\in \mathbb{H}^{N_{\text{\tiny{T}}}}, \mathbf{\widetilde Q}_{k}\in \mathbb{H}^{N_{\tiny \mathrm{U}_k}},\\ {\tau_k}, \omega_k, \theta_k}}{\maxo}\,\, \hspace*{-2mm}&& \sum_{k=1}^K  \min_{\norm{\Delta \mathbf{H}_k}_2\le \rho_k}\, \tau_k\log_2\,\det\Big(\mathbf{I}_{\mathrm{R}}+\frac{1}{\sigma_n^2}\mathbf{H}_k^H\frac{\mathbf{\widetilde Q}_k}{\tau_k}\mathbf{H}_k\Big)\label{eqn:TP_maximization_20a}\\
\hspace*{-2mm}\mathrm{subject\,\,to}\,\, &&\mathrm{C1 - C7}.\notag
 \end{eqnarray}
 
 Since the original problem in \eqref{eqn:TP_maximization} with the objective function defined with respect to the $\norm{\cdot}_\mathrm{F}$-norm cannot be efficiently solved, in the following, we aim at finding an optimal solution of problem \eqref{eqn:TP_maximization_20a} with the objective function defined with respect to the spectral norm. 
However, before we are able to solve the transformed problem \eqref{eqn:TP_maximization_20a} efficiently, we have to handle the remaining difficulty in dealing with the CSI uncertainty in the objective function.
\begin{Lem}[Theorem 1 \cite{JR:Robust_MIMO_uncertainty}]
Let $\widetilde{\mathbf{Q}}_{k}$ $\in$ $\mathcal{Q}$, where $\mathcal{Q}$ is a nonempty compact convex set that satisfies
\begin{eqnarray}\label{eqn:uni_inv}
&&\mathbf{U}_k \widetilde{\mathbf{Q}}_{k} \mathbf{U}_k \in \mathcal{Q},\label{eqn:uni_inv_1}\\
&&\mathbf{D}(\widetilde{\mathbf{Q}}_{k}) \in \mathcal{Q}, \forall k,\label{eqn:uni_inv_2}
\end{eqnarray}
for all $\widetilde{\mathbf{Q}}_{k}$ $\in \mathcal{Q}$ and all unitary matrices $\mathbf{U}_k \in \mathbb{C}^{N_{\mathrm{U}_k} \times N_{\mathrm{U}_k}}$, where $\mathbf{D}(\widetilde{\mathbf{Q}}_{k})$ is a diagonal matrix having the same
diagonal elements as $\widetilde{\mathbf{Q}}_{k}$. Moreover, let $\Delta \mathbf{H}_k = \mathbf{H}_k -  \mathbf{\widehat H}_k$ $\in$ $\mathbf{\widetilde \Lambda}_{k}$, where
\begin{equation}
 \mathbf{\widetilde \Lambda}_{k} = \Big\{\Delta\mathbf{H}_k\in \mathbb{C}^{N_{\mathrm{U}_k}\times {N_{\mathrm{R}}}}:\norm{\Delta\mathbf{H}_k}_2 \le \rho_k\Big\}, \forall k.
 \end{equation} Then, the optimal $\widetilde{\mathbf{Q}}_{k}$ in the optimization problem in \eqref{eqn:TP_maximization_20a}, denoted by $\mathbf{\widetilde Q}_k^*$, has the following form:
 \begin{equation}
  \mathbf{\widetilde Q}_k^* = \mathbf{V}_{0k} \mathbf{\Lambda}_{\widetilde{\mathbf{ Q}}_k^*} \mathbf{V}_{0k}^H, \forall k,
  \end{equation}
  where $\mathbf{\Lambda}_{\widetilde{\mathbf{ Q}}_k^*} = \mathrm{diag}(\bm{\widetilde \lambda}_k^*)$, $\widetilde{\bm{\lambda}}_k^* = \{ \widetilde{\lambda}_{i, k}^*\}$, $i = \{1, \ldots, \mathrm{min}\{N_{\mathrm{U}_k},N_\mathrm{R}\}\}$, $k = \{1,\ldots,K\},$ contains the eigenvalues of the optimal transmit covariance matrix $\widetilde{\mathbf{Q}}_k^*$ at wireless powered user $k$, and $\mathbf{V}_{0k} $ is a unitary matrix obtained from the singular value decomposition (SVD) of the estimated channel $\widehat{\mathbf{H}}_k = \mathbf{U}_{0k} \mathbf{\Sigma}_{\widehat{\mathbf{H}}_k} \mathbf{V}_{0k}$, $\forall k$.
The optimum solution $\widetilde{\bm{\lambda}}_k^*$ can be obtained by solving the following optimization problem 
  \begin{equation}\label{eqn:updated_problem_2}
 \underset{ \mathbf{V}\in \mathbb{H}^{N_\mathrm{T}}, \bm{\widetilde \lambda}_k,\bm{\gamma}_k, {\tau_k}, \omega_k, \theta_k}{\maxo}\,\, \hspace*{-2mm} \sum_{k=1}^K  \min_{\norm{\bm{\gamma}_k - \bm{\widehat \gamma}_{k}}_{\infty}\le \rho_k}\, \sum_{i=1}^{\mathrm{min}\{N_{\mathrm{U}_k},N_\mathrm{R}\}} \tau_k\log_2\,\bigg(1+\frac{\gamma_{i,k}^2}{\sigma_n^2 \tau_k}{\widetilde \lambda_{i,k}}\bigg),
 \end{equation}
 where the constraint set is identical to that in \eqref{eqn:TP_maximization_20a}. The auxiliary optimization variables $\bm{\gamma}_{k} = \{\gamma_{i,k} \}, \forall i, k$, represent the singular values of channel matrix $\mathbf{H}_k$, and $\bm{\hat \gamma}_{k} = \{\widehat{\gamma}_{i,k} \}, \forall i, k$, are the singular values of the estimated channel matrix $\widehat{\mathbf{H}}_k.$\end{Lem}
  \emph{Proof: }Please refer to \cite[Appendix A]{JR:Robust_MIMO_uncertainty} for a proof of Lemma 2.\qed

  Lemma 2 states that if the covariance matrix $\mathbf{\widetilde Q}_k$ belongs to a set $\mathcal{Q}$ that satisfies the unitarily invariant set properties in \eqref{eqn:uni_inv_1} and \eqref{eqn:uni_inv_2}, the optimal solution of \eqref{eqn:TP_maximization_20a} can be obtained by solving an equivalent optimization problem with the objective function given in \eqref{eqn:updated_problem_2}. In fact, in problem \eqref{eqn:TP_maximization_20a}, constraints C3a and C6 describe an unitarily invariant sum power constraint set for $\mathbf{\widetilde Q}_k$, which satisfies \eqref{eqn:uni_inv_1} and \eqref{eqn:uni_inv_2} \cite{JR:Robust_MIMO_uncertainty}. Thus, in the sequel, we adopt the objective function in \eqref{eqn:updated_problem_2} for the development of the proposed resource allocation algorithm. Next, we tackle the auxiliary optimization variables $\bm{ \gamma}_k$ in the following lemma.
\begin{Lem}[Theorem 3 \cite{JR:Robust_MIMO_uncertainty}]	If the conditions of Lemma 2 are fulfilled, such that $\mathbf{\widetilde Q}_k$ $\in \mathcal{Q}$, $\forall k,$ where $\mathcal{Q}$ satisfies the unitarily invariant set properties in \eqref{eqn:uni_inv_1} and \eqref{eqn:uni_inv_2}, and $\Delta \mathbf{H}_k \in$ $\mathbf{\widetilde \Lambda}_{k}$, then the solution for the singular values of the worst possible channel $\mathbf{H}_k, \forall k$, in problem \eqref{eqn:updated_problem_2} is given by
 \begin{equation}\label{opt_solution}
\gamma_{i,k}^* = \mathrm{max} \{\widehat{\gamma}_{i,k}-\rho_k,0\}, \, \, \forall k, i \in \{1,\ldots, \min \{N_{\mathrm{U}_k},N_\mathrm{R}\}\}.
\end{equation}\end{Lem}
 \emph{Proof: }Please refer to \cite[Theorem 3]{JR:Robust_MIMO_uncertainty} for a proof of Lemma 3.\qed
 
Lemma 3 states that the optimal solution for the sum throughput optimization problem in \eqref{eqn:TP_maximization_20a} with the objective function given in \eqref{eqn:updated_problem_2} diagonalizes the estimated channel $\widehat{\mathbf{H}}_k, \forall k$, via SVD and exploits its singular values $\widehat{\gamma}_{i,k}, \forall i,k$, as in \eqref{opt_solution}.

Applying the results from Lemma 2 and Lemma 3, we obtain the following equivalent simpler optimization problem:
 \begin{eqnarray}\label{eqn:updated_problem_final}
 \underset{ \mathbf{V}\in \mathbb{H}^{N_{\text{\tiny{T}}}}, \bm{\widetilde \lambda}_k,{\tau_k}, \omega_k, \theta_k}{\maxo}\,\, \hspace*{-2mm}&& \sum_{k=1}^K  \, \sum_{i=1}^{\mathrm{min}\{ \tiny N_{\mathrm{U}_k},{N_{\text{\tiny{R}}}}\}} \tau_k\log_2\,\bigg(1+\frac{\widetilde{\lambda}_{i, k}}{\sigma_n^2 \tau_k} \mathrm{max}\{\widehat{\gamma}_{i, k}-\rho_k, 0 \}^2\bigg)\\
\hspace*{-2mm}\mathrm{subject\,\,to}\,\, &&\mathrm{C1, C2},\notag\\
\hspace*{-2mm}&&\mathrm{C3a}:\,\, T_{\max} P_{\mathrm{c}_k}+\sum_{i=1}^{\mathrm{min}\{ N_{\tiny \mathrm{U}_k},{N_{\text{\tiny{R}}}}\}} \widetilde{\lambda}_{i, k}\varepsilon_k\leq \tau_0 \frac{\frac{M_k}{1+\exp\big(\hspace*{-0.5mm}-a_k(\theta_k-\hspace*{-0.5mm}b_k)\big)}
 \hspace*{-0.5mm}- \hspace*{-0.5mm}M_k\Omega_k}{1-\Omega_k} ,\,\forall k,\notag\\
\hspace*{-2mm}&&\mathrm{C3b, C4, C5, C7},\,\, \mathrm{C6}:\,\, \widetilde{\lambda}_{i, k} \geq 0,\forall i, k.\notag
  \notag \end{eqnarray}

\subsection{Dual Problem Formulation and Solution}

It can be shown that, for a given $\tau_0$, problem \eqref{eqn:updated_problem_final} is a convex optimization problem and satisfies Slater's constraint qualification. Thus,
strong duality holds, the duality gap is equal to zero, and solving the dual problem is equivalent to solving the primal problem \cite{book:convex}. In order to reveal the structure of the solution and to obtain some system design insight, in the following, we study the dual solution of problem \eqref{eqn:updated_problem_final}. To obtain the dual solution, we first need the Lagrangian function for problem \eqref{eqn:updated_problem_final}, which is given by:
 \begin{eqnarray}
\mathcal{L} && =  \sum_{k=1}^K \sum_{i=1}^{\mathrm{min}\{N_{\tiny \mathrm{U}_k},N_{\text{\tiny{R}}}\}} \tau_k\log_2\,\bigg(1+\frac{\widetilde{\lambda}_{i, k}}{\sigma_n^2 \tau_k} \mathrm{max}\{\widehat{\gamma}_{i, k}-\rho_k, 0 \}^2\bigg)\label{eqn:lagrangian}\\
&&- \mu \big(\Tr(\mathbf{V}) - P_{\max}\big) - \kappa \bigg(\sum_{k=0}^K \tau_k - T_{\max}\bigg)\notag\\
&&-\sum_{k=1}^K \beta_k \Bigg[ T_{\max} P_{\mathrm{c}_k} +  \sum_{i=1}^{\mathrm{min}\{N_{\tiny \mathrm{U}_k},N_{\text{\tiny{R}}}\}} \widetilde{\lambda}_{i, k}\varepsilon_k - \tau_0 \frac{\frac{M_k}{1+\exp\big(\hspace*{-0.5mm}-a_k(\theta_k-\hspace*{-0.5mm}b_k)\big)} - M_k \Omega_k}{1-\Omega_k}\Bigg]\notag\\
 &&+\sum_{k=1}^K \Tr\big(\mathbf{\Upsilon} (\bm{\mathcal V}, \omega_k, \theta_k\big) \mathbf{M}_{\mathrm{C3b}_k}) + \Tr(\mathbf{V}\mathbf{M}_{\mathrm{C5}})+\sum_{k=1}^K \sum_{i=1}^{\mathrm{min}\{N_{\tiny \mathrm{U}_k},N_{\text{\tiny{R}}}\}} \xi_{i, k} \widetilde{\lambda}_{i, k}.\notag
\end{eqnarray}
In \eqref{eqn:lagrangian}, $\mathbf{M}_{\mathrm{C3b}_k} \succeq \bm{0}$ and $\mathbf{M}_{\mathrm{C5}} \succeq \bm{0}$ are the Lagrange multiplier matrices corresponding to constraints C3b and C5, respectively. $\mu \geq 0$ is the Lagrange multiplier that accounts for the total transmit power constraint C1. Also, $\kappa \geq 0$ is the Lagrange multiplier related to the total time constraint in C2. $\beta_k \geq 0, \forall k,$ and $\xi_{i,k} \geq 0, \forall i, k, $ account for the total power consumption constraint in C3a and C6, respectively. The dual problem of problem \eqref{eqn:updated_problem_final} is given by:
\begin{equation}\label{eqn:dual_problem}
\underset{\substack{\mathbf{M}_{\mathrm{C3b}_k}, \mathbf{M}_{\mathrm{C5}}, \\ \mu, \kappa,\beta_k,\xi_{i, k}}}{\mathrm{minimize}} \,\,\, \underset{\substack{\mathbf{V}\in \mathbb{H}^{N_{\text{\tiny{T}}}}, \bm{\widetilde \lambda}_k,{\tau_k},\\ \omega_k, \theta_k}}{\maxo} \mathcal{L}.
\end{equation}

For optimization problems that satisfy Slater's constraint qualification, the Karush-Kuhn-Tucker (KKT) conditions are necessary and sufficient conditions for the solution of the problem \cite{book:convex}. The KKT conditions for \eqref{eqn:updated_problem_final}, with respect to the optimal solution for $\mathbf{V}^*$, $\tau_k^*$, and $\widetilde{\lambda}_{i,k}^*, \forall i, k$, are as follows:
\begin{subequations}
\begin{eqnarray}\label{KKT_conditions}
\mathbf{M}_{\mathrm{C3b}_k}^*, \mathbf{M}_{\mathrm{C5}}^* \succeq \mathbf{0}, \,\, \mu^*, \kappa^*, \beta_k^*, \xi_{i,k}^* &\geq& 0, \forall i, k, \label{subseq:eq1}\\
\mu^*\Big( \Tr(\mathbf{V}^*) - P_{\mathrm{max}}\Big) &=& 0, \label{subseq:eq1_1}\\
\kappa^*\Big(\sum_{k=0}^{K}\tau_k - T_{\mathrm{max}}\Big) &=& 0, \label{subseq:eq1_2}\\
\beta_k^* \Big( T_{\max} P_{\mathrm{c}_k} +  \sum_{i=1}^{\mathrm{min}\{N_{\tiny \mathrm{U}_k},N_{\text{\tiny{R}}}\}} \widetilde{\lambda}_{i, k}^* \varepsilon_k - \tau_0^* \Phi_k^{\mathrm{Practical}}(\theta_k^*)\Big) &=& 0, \forall k, \label{subseq:eq2}\\
\mathbf{\Upsilon} (\bm{\mathcal V}^*, \omega_k^*, \theta_k^*) \mathbf{M}_{\mathrm{C3b}_k}^* &=& \mathbf{0}, \forall k,  \label{subseq:eq2_1} \\
\mathbf{V}^* \mathbf{M}_{\mathrm{C5}}^* = \mathbf{0},\, \xi_{i, k}^* \widetilde{\lambda}_{i,k}^* &=& 0, \forall i, k,\label{subseq:eq3} \\
\nabla_{\mathbf{V}^*} \mathcal{L} = \mathbf{0},\,\,
\frac{\partial \mathcal{L}}{\partial \widetilde{\lambda}_{i,k}^*} = 0,\, \frac{\partial \mathcal{L}}{\partial \tau_{k}^*} &=& 0,  \forall i, k.\label{subseq:eq6}
\end{eqnarray}
\end{subequations}
Here,  $\mathbf{M}_{\mathrm{C3b}_k}^*$, $\mathbf{M}_{\mathrm{C5}}^*$, $\mu^*$, $\beta_k^*$, and $\xi_{i,k}^*$, $\forall i, k$, are the optimal Lagrange multipliers for the dual problem in \eqref{eqn:dual_problem}. Moreover, $\Phi_k^{\mathrm{Practical}}(\theta_k^*)$ denotes the harvested power based on the non-linear EH model in \eqref{eqn:EH_non_linear} for the optimal received power $\theta_k^*$.

In the following theorem, we investigate the optimal structure of the energy matrix.
\begin{Thm}\label{sdp_relaxation}
If problem \eqref{eqn:updated_problem_final} is feasible and $P_{\max}>0$, the optimal energy matrix $\mathbf{V}$ is a rank-one matrix and can be expressed as
\begin{equation}\label{rank_one_structure}
\mathbf{V}^* = P_{{\max}}\mathbf{u}_{\mathbf{\Gamma},\max}\mathbf{u}_{\mathbf{\Gamma},\max}^H,
\end{equation}
where $\mathbf{u}_{\mathbf{\Gamma},\max}\in \mathbb{C}^{N_{\mathrm{T}}\times 1}$ is the unit-norm eigenvector corresponding to the maximum eigenvalue of matrix $\mathbf{\Gamma} \triangleq \sum_{k=1}^{K} \sum_{l=1}^{N_{\tiny \mathrm{U}_k} }\Big[\mathbf{U}_{\widehat{\mathbf{g}}_k} \mathbf{M}_{\mathrm{C3b}_k}^*\mathbf{U}_{\widehat{\mathbf{g}}_k}^H\Big]_{a:b,c:d}$ with $a = (l-1)N_{\mathrm{T}} + 1, b=l N_{\mathrm{T}}, c= (l-1)N_{\mathrm{T}} + 1 $, $d=l N_{\mathrm{T}}$.
\end{Thm}
\emph{Proof: } Please refer to Appendix A.

Theorem \ref{sdp_relaxation} reveals that the optimal solution for the energy matrix,  $\mathbf{V}^*$, is a rank-one matrix and thus beamforming is optimal for the maximization of the sum throughput in the MIMO-WPCN, despite the CSI uncertainty and the non-linear EH model. In particular, the beamforming direction, i.e., $\mathbf{u}_{\mathbf{\Gamma},\max}$, is aligned with the maximum eigenmode of matrix $\mathbf{\Gamma}$, which depends on the estimated downlink channel matrix $\widehat{\mathbf{G}}_k$.

On the other hand, exploiting the fact that the partial derivative of the Lagrangian with respect to $\widetilde{\lambda}_{i, k}^*$ vanishes at the optimal solution, from \eqref{subseq:eq6}, we obtain:
\begin{equation}\label{eqn:eig_val_solution}
\lambda_{i, k}^* = \frac{\widetilde{\lambda}_{i, k}^*}{\tau_k^*}  = \Bigg[ \frac{1}{\ln(2)(\beta_k^*\varepsilon_k)} - \frac{\sigma_n^2}{\gamma_{i,k}^*} \Bigg]^{+}, \, \, \forall i, k,
\end{equation}
where $x^{+} = \max \{ 0, x \}$. Eq. \eqref{eqn:eig_val_solution} reveals that the optimal power allocation for the eigenmodes of the precoding matrix has a water-filling structure. The dual variable $\beta_k^*$ in \eqref{eqn:eig_val_solution} ensures that considering the power harvested during the downlink WET period $\tau_0^*$, the individual power consumption constraint\footnote{We note that it can be shown that according to the optimal resource allocation solution, if $\tau_k^* \neq 0$, then wireless powered user $k$ exhausts all of the power harvested during the WET period in order to facilitate its information transfer in the WIT period and to maximize the system objective. The proof is omitted due to page limitation. } at wireless powered user $k$, $\forall k$, is satisfied. Furthermore, wireless powered users with better channel conditions and more reliable channel estimates are allocated more power, as their value of $\gamma_{i,k}^*$ is larger, cf. \eqref{opt_solution}. We note that the values for the dual variables $\beta_k^*, \forall k,$ used for the calculation of $\lambda_{i,k}^*$, can be obtained via various algorithms, such as the  subgradient method or the ellipsoid method \cite{book:convex}.

In the following proposition, we study the optimal solution for the time allocation for WET and WIT, $\tau_0^*$ and $\tau_k^*, k \in \{1, \ldots, K\},$ respectively.
\begin{proposition}\label{time_allocation_proposition}
The optimal time allocation solution for problem \eqref{eqn:updated_problem_final} is given by
\begin{eqnarray}
\tau_0^* &=& T_{\max} \frac{1+\sum_{k=1}^{K}\frac{P_{\mathrm{c}_k}}{\sum_{i=1}^{\mathrm{min}\{N_{\tiny \mathrm{U}_k},N_\mathrm{R}\}}\lambda_{i, k}^*\varepsilon_k}}{1 + \sum_{k=1}^{K} \frac{\Phi_k^{\mathrm{Practical}}(\theta_k^*)}{\sum_{i=1}^{\mathrm{min}\{N_{\tiny \mathrm{U}_k},N_\mathrm{R}\}}\lambda_{i, k}^*\varepsilon_k}}, \label{eqn:tau0prop}\\
\tau_k^* &=& \frac{\tau_0^* \Phi_k^{\mathrm{Practical}}(\theta_k^*) - T_{\max}P_{\mathrm{c}_k}}{\sum_{i=1}^{\mathrm{min}\{N_{\tiny \mathrm{U}_k},N_\mathrm{R}\}} \lambda_{i, k}^*\varepsilon_k}, \,\, \forall k = \{ 1, \ldots, K\}, \label{eqn:tauKprop}
\end{eqnarray}
where $\lambda_{i, k}^*, \forall i,k,$ is given by \eqref{eqn:eig_val_solution}.
\end{proposition}

\emph{Proof: } Please refer to Appendix B.

Proposition \ref{time_allocation_proposition} provides an analytical solution for the optimal time allocation in problem \eqref{eqn:updated_problem_final} and significant insights for system design. 
For example, for increasing channel estimation errors for the channel between the power station and the wireless powered users, $\lambda_{i, k}^*$ decreases, cf. \eqref{opt_solution}, \eqref{eqn:eig_val_solution}, and since $P_{\mathrm{c}_k} \leq  \Phi_k^{\mathrm{Practical}}(\theta_k^*)$ always holds, cf. C2, C3a, the WET duration $\tau_0^*$ increases, cf. \eqref{eqn:tau0prop}. Hence, a longer WET period is needed when the CSI uncertainty increases. On the other hand, similar considerations for $\tau_k^*$, after substituting \eqref{eqn:tau0prop} into \eqref{eqn:tauKprop}, reveal that the WIT period $\tau_k^*$ decreases when the CSI uncertainty increases.

\subsection{Solution of the Max-min Optimization Problem}

In order to solve the optimization problem in \eqref{eqn:TP_max_min}, we first employ the same transformation as in Sections III.A and III.B. Additionally, we introduce an auxiliary optimization variable $\nu$ that denotes the minimum throughput achieved by each individual wireless powered user. Then, by exploiting Lemma 2 and Lemma 3, problem \eqref{eqn:TP_max_min} is transformed into the following equivalent convex optimization problem:
\begin{eqnarray}\label{eqn:updated_problem_final_max_min}
&& \underset{ \mathbf{V}\in \mathbb{H}^{N_{\text{\tiny{T}}}}, \bm{\widetilde \lambda}_k,{\tau_k}, \omega_k, \theta_k, \nu}{\maxo}\,\, \nu \\
\hspace*{-2mm}\mathrm{subject\,\,to}\,\,&& \mathrm{C1}:\,\,\Tr(\mathbf{V})\le P_{\max}, \, \, \mathrm{C2}:\,\, \tau_0 +\sum_{k=1}^K \tau_k \leq T_{\max},\notag\\
\hspace*{-2mm}&&\mathrm{C3a}:\,\, T_{\max} P_{\mathrm{c}_k}+\sum_{i=1}^{\min \{N_{\tiny \mathrm{U}_k}, N_{\text{\tiny{R}}}\}} \widetilde{\lambda}_{i, k}\varepsilon_k\leq \tau_0 \frac{\frac{M_k}{1+\exp\big(\hspace*{-0.5mm}-a_k(\theta_k-\hspace*{-0.5mm}b_k)\big)}
 \hspace*{-0.5mm}- \hspace*{-0.5mm}M_k\Omega_k}{1-\Omega_k} ,\,\forall k,\notag\\
\hspace*{-2mm}&&\mathrm{C3b}:\,\, \mathbf{\Upsilon} (\bm{\mathcal{V}}, \omega_k, \theta_k) \succeq \mathbf{0} ,\,\forall k,\notag\\
\hspace*{-2mm}&&\mathrm{C4}\,\, ,\mathrm{C5} \,\, ,\mathrm{C7}, \,\,\mathrm{C6}:\,\, \widetilde{\lambda}_{i, k} \geq 0,\forall i, k,  \notag\\
\hspace*{-2mm}&&\mathrm{C8}:\,\, \tau_k\log_2\,\big(1+\frac{\widetilde{\lambda}_{i, k}}{\sigma_n^2 \tau_k} \mathrm{max}\{\widehat{\gamma}_{i, k}-\rho_k, 0 \}^2\big) \geq \nu,\forall k.  \notag
 \end{eqnarray} 
 The dual of problem \eqref{eqn:updated_problem_final_max_min} can be obtained in a similar manner as the dual problem given in \eqref{eqn:dual_problem}. Moreover, similar as in \eqref{eqn:eig_val_solution}, we can obtain the power allocation for the max-min scheme as
\begin{equation}\label{eqn:eig_val_solution_max_min}
\lambda_{i, k}^* = \Bigg[ \frac{\varsigma_k^*}{\ln(2)(\beta_k^*\varepsilon_k)} - \frac{\sigma_n^2}{\gamma_{i,k}^*} \Bigg]^{+}, \, \, \forall i, k,
\end{equation}
where $\varsigma_k^*$  is the optimal Lagrange multiplier associated with C8 in \eqref{eqn:updated_problem_final_max_min} which accounts for the individual throughput of each wireless powered user $k$. Unlike the power allocation in \eqref{eqn:eig_val_solution}, the additional parameter $\varsigma_k^*$ in \eqref{eqn:eig_val_solution_max_min} ensures fairness among different wireless powered users. The optimal time allocation is still given by \eqref{eqn:tau0prop} and \eqref{eqn:tauKprop}, but with $\lambda_{i, k}^*$ from \eqref{eqn:eig_val_solution_max_min}.
  In addition, exploiting the solution of the dual problem of  \eqref{eqn:updated_problem_final_max_min} and Theorem \ref{sdp_relaxation}, it can be shown that the optimal energy matrix $\mathbf{V}^*$ still has the structure given in \eqref{rank_one_structure}.

\begin{table}[t]\caption{Resource Allocation Algorithm.}\label{table:algorithm}
\vspace*{-5mm}
\begin{algorithm} [H]                    
\caption{Proposed Resource Allocation Algorithm}          
\label{alg1}                           
\begin{algorithmic}[1]
\STATE {Initialize
$\gamma_{i,k}^*, \forall i, k$, according to \eqref{opt_solution} and  $\tau_0$}
\REPEAT[Outer Loop]
\STATE{Initialize the maximum number of iterations $I_{\text{max}}$, iteration index $m = 0$, and $\theta_k^{(0)}, \forall k$}
\REPEAT[Successive Convex Approximation]
\STATE{Solve the max-sum/max-min optimization problem in \eqref{eqn:updated_problem_final}/\eqref{eqn:updated_problem_final_max_min}, with $\Phi_{\text{EH}_k}^{\text{Practical}} (\theta_k') = \Phi_{\text{EH}_k}^{\text{Practical}} (\theta_k^{(m)}) + \nabla_{\theta_k} \Phi_{\text{EH}_k}^{\text{Practical}} (\theta_k^{(m)}) \big(\theta_k' - \theta_k^{(m)} \big)$ and obtain the intermediate $\theta_k', \forall k$}
\STATE{Update $\theta_{k}^{(m+1)} = \theta_k', \forall k, \, m=m+1$}
\UNTIL{Convergence or $m = I_{\text{max}}$}
\STATE{Obtain the solution for $\mathbf{V}$, $\tau_k$, $\tilde{\bm{\lambda}}_k$, $\forall k$}
\STATE {Increase $\tau_0$}
\UNTIL{$\tau_0 = T_{\mathrm{max}}$}
\STATE {Perform a one-dimensional search to find the optimal $\tau_0^*$ that yields the maximum objective value}
\STATE{Obtain $\tau_k^*$, $\widetilde{\mathbf{Q}}_k^* = \mathbf{V}_{0k} \mathbf{\Lambda}_{\mathbf{\widetilde{Q}}_k^*}\mathbf{V}_{0k}^H$, and $\mathbf{V}^*$, where $\mathbf{\Lambda}_{\mathbf{\widetilde{Q}}_k^*} = \mathrm{diag}(\bm{\widetilde \lambda}_{k}^*)$, $\forall k$}
\end{algorithmic}
\end{algorithm}
\vspace*{-8mm}
\end{table}

\subsection{Overall Resource Allocation Algorithm}

In the following, we present the overall resource allocation algorithm  for achieving the globally optimal solution of the max-sum resource allocation problem in \eqref{eqn:updated_problem_final} and the max-min resource allocation problem in \eqref{eqn:updated_problem_final_max_min}. The structure of the algorithm is depicted in Table \ref{table:algorithm}. The considered optimization problems are solved for a given value of $\tau_0$, by exploiting numerical solvers for convex programs such as CVX \cite{website:CVX}. Then, we perform a one-dimensional search to obtain the optimal $\tau_0$ in order to obtain the optimal variables $\tau_k^*$, $\widetilde{\mathbf{Q}}_k^*$, and $\mathbf{V}^*$, as specified in Table \ref{table:algorithm}. 

\emph{Remark 1: }When the non-linear EH model is adopted for resource allocation, some practical implementation issues may arise. More specifically, some popular numerical convex program solvers, such as CVX \cite{website:CVX}, are not able to directly handle constraint C3a in \eqref{eqn:updated_problem_final} and \eqref{eqn:updated_problem_final_max_min}, even though the constraint is convex for a given $\tau_0$. To overcome this problem, we adopt the successive convex approximation method \cite{book:convex} such that existing numerical solvers can be employed. Specifically, for a given $\tau_0,$ the right-hand side of constraint C3a is a differentiable concave function. Then, the following inequality always holds for any feasible point $\theta_k^{(m)}$:
\begin{equation}\label{eqn:succ_convex_app}
\Phi_{\text{EH}_k}^{\text{Practical}} (\theta_k) \leq \Phi_{\text{EH}_k}^{\text{Practical}} (\theta_k^{(m)}) + \nabla_{\theta_k} \Phi_{\text{EH}_k}^{\text{Practical}} (\theta_k^{(m)}) \big(\theta_k - \theta_k^{(m)} \big), 
\end{equation}
where $m$ denotes the iteration index. It follows that, for a given $\theta_k^{(m)}$, solving the optimization problems in \eqref{eqn:updated_problem_final}/\eqref{eqn:updated_problem_final_max_min}  after replacing the right-hand side of constraint C3a with \eqref{eqn:succ_convex_app}, leads to an upper bound on the optimal values of these optimization problems. In order to tighten this upper bound, we use an iterative algorithm, which starts with the initialization of the value of $\theta_k^{(m)}$ and iteration index $m=0$, cf. lines 3 - 7 in Algorithm \ref{alg1}. We obtain an intermediate solution $\theta_k', \forall k$, by solving the convex optimization problem in \eqref{eqn:updated_problem_final}/\eqref{eqn:updated_problem_final_max_min} with $\Phi_{\text{EH}_k}^{\text{Practical}} (\theta_k') = \Phi_{\text{EH}_k}^{\text{Practical}} (\theta_k^{(m)}) + \nabla_{\theta_k} \Phi_{\text{EH}_k}^{\text{Practical}} (\theta_k^{(m)}) \big(\theta_k' - \theta_k^{(m)} \big)$, cf. \eqref{eqn:succ_convex_app}. Then, we update $\theta_k^{(m+1)}$ with $\theta_k', \forall k.$ These steps are repeated iteratively until convergence or the maximum number of iterations are reached. We note that using this technique, since the right-hand side of constraint C3a is a concave function, the iterative algorithm, successive convex approximation, always converges to the optimum solution with respect to the original problem formulation in \eqref{eqn:updated_problem_final}/\eqref{eqn:updated_problem_final_max_min} with polynomial time computational complexity \cite{book:convex}. Specifically, from our simulation experience, we found that less than 5 iterations were required for convergence of the proposed resource allocation algorithm for each channel realization.

\section{Numerical Results}\label{sect:simulation}

\begin{table}[t]\vspace*{-4mm}
\caption{Simulation Parameters.} \label{table:parameters}
\centering
\begin{tabular}{ | l | l | } \hline
      Carrier center frequency                           & 915 MHz\\ \hline
      Bandwidth                                          & $200$ kHz \\ \hline
      Path loss exponent 								 & $3.6$ \\ \hline
            Power station-to-wireless powered users fading distribution                                      & Rician with Rician factor $3$ dB \\ \hline
      Wireless powered users-to-information receiving station fading distribution                                      & Rayleigh \\
        \hline
      Power station  antenna gain                         & $10$ dBi \\ \hline
      Information receiving station  antenna gain        & $2$ dBi \\ \hline
      Noise power                                        & $\sigma_n^2= -95$ dBm \\ \hline
      Power amplifier efficiency						 & $20\% $ \\ \hline
      Circuit power consumption							 & $P_{\mathrm{c}_k} = 5 \mu$W, $\forall k$ \\ \hline

\end{tabular}\vspace*{-4mm}
\end{table}

In this section, we evaluate the system performance, for both the proposed max-sum and the  proposed max-min resource allocation algorithms. The simulation parameters are provided in Table \ref{table:parameters}. For the wireless channel, we adopt the TGn path loss model \cite{report:tgn}. We assume that the power station and the information receiving station in the considered MIMO-WPCN are equipped with $N_{\mathrm{T}} = 4$ transmit and $N_{\mathrm{R}} = 4$ receive antennas, respectively, unless specified otherwise. Each wireless powered user is assumed to have 2 antennas, i.e., $N_{\mathrm{U}_k} = N_{\text{U}} = 2, \forall k$. The wireless powered users are randomly and uniformly distributed between the reference distance of $2$ meters and the maximum WET service distance of $20$ meters from the power station. The information receiving station is $100$ meters away from the power station. We assume $K = 4$ wireless powered users in the MIMO-WPCN, unless indicated differently. Regarding the non-linear EH model parameters, cf. \eqref{eqn:EH_non_linear}, we assume $M_k = 24$ mW, $b_k = 0.0022$, and $a_k = 1500$, $\forall k$ \cite{CN:EH_measurement_2}. For simplicity, we normalize the duration of a communication slot to $T_{\mathrm{max}} = 1$. Moreover, we define the normalized maximum channel estimation error $\sigma_{\mathrm{est}_k}^2 $ such that $\sigma_{\mathrm{est}_k}^2 \ge \upsilon_k^2 / \norm{\mathbf{G}_k}_2^2$ and $\sigma_{\mathrm{est}}^2 \ge \rho_k^2 / \norm{\mathbf{H}_k}_2^2$, $\forall k$, and we assume that $\sigma_{\mathrm{est}}^2 $ is identical for all wireless powered users.  For the proposed resource allocation algorithm,  we  quantize the possible range of $\tau_0$, $0 \leq \tau_0 \leq T_{\mathrm{max}}$, into  $50$  equally  spaced  intervals for  conducting  the  full  search. The results obtained in this section were averaged over 1000 path loss and small scale fading realizations.

\begin{figure}[t]
\centering
\includegraphics[width=3.4 in]{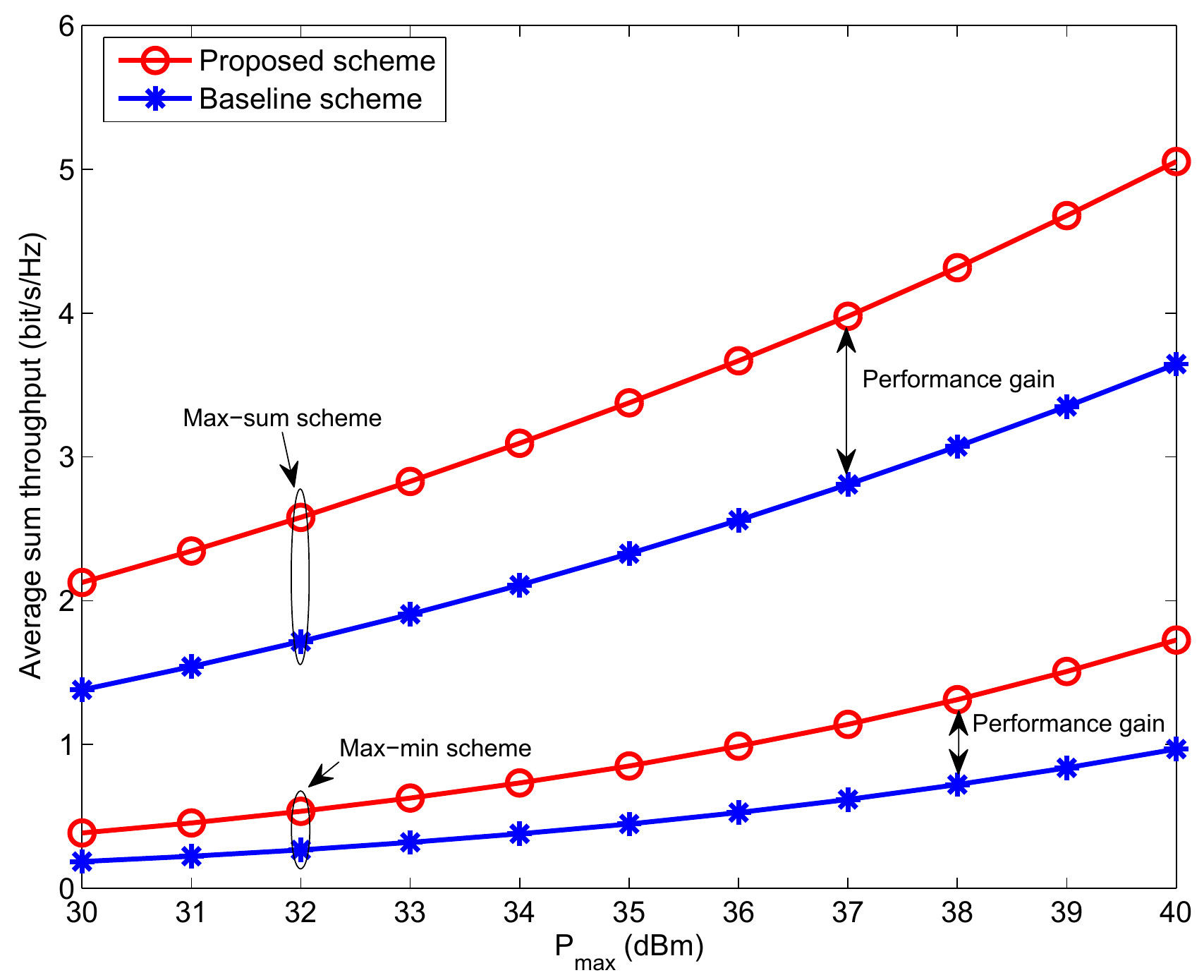}
\caption{Average sum throughput (bit/s/Hz) versus the maximum transmit power at the power station $P_{\max }$ (dBm) for $K=4$ wireless powered users and $\sigma_{\mathrm{est}}^2 = 0.05$.}\vspace*{-5mm}
\label{fig_1:av_sum_throughput_vs_pmax_4_users}
\end{figure}
Figure \ref{fig_1:av_sum_throughput_vs_pmax_4_users} depicts the average sum throughput versus the maximum transmit power allowance $P_{\max }$ for $K = 4$ wireless powered users, and a normalized maximum channel estimation error of $\sigma_{\mathrm{est}}^2 = 5\%$.  The average sum throughput is depicted for both the  max-sum resource allocation scheme and the max-min resource allocation scheme, cf. \eqref{eqn:updated_problem_final} and \eqref{eqn:updated_problem_final_max_min}, respectively. As can be observed, for all considered schemes, the sum throughput is monotonically increasing with respect to the maximum transmit power at the power station $P_{\max }$. This is due to the fact that for a given $\tau_0$, the wireless powered users can harvest more energy for WIT for a larger value of $P_{\max }$. Furthermore, the proposed max-sum resource allocation scheme achieves the highest sum throughput. In contrast, the max-min scheme tries to balance the throughput achieved by all wireless powered users, which is at the expense of a poorer sum throughput performance, since the channels of different wireless powered users may differ significantly. 
For comparison, we also show the performance of a baseline scheme, which performs the resource allocation based on the linear EH model in \eqref{eqn:linear_model} subject to the constraint set in \eqref{eqn:updated_problem_final}. The power conversion efficiency for the linear EH model is selected as $\eta_k = 0.5, \forall k$ \cite{JR:WPC_Rui_Zhang}. As can be observed, the results for the baseline schemes show a performance degradation compared to the proposed schemes. In particular, the RF power directed at the wireless powered users for the resource allocation scheme based on the linear EH model may cause saturation at the EH receivers of some wireless powered users and underutilization of other wireless powered users since the linear EH model does not account for the non-linearity of practical EH circuits. This leads to resource allocation mismatches, which result in a poor performance for the baseline scheme for both system design objectives.

\begin{figure}
\centering
\begin{minipage}{0.5\linewidth}
\centering
\includegraphics[width=1\linewidth]{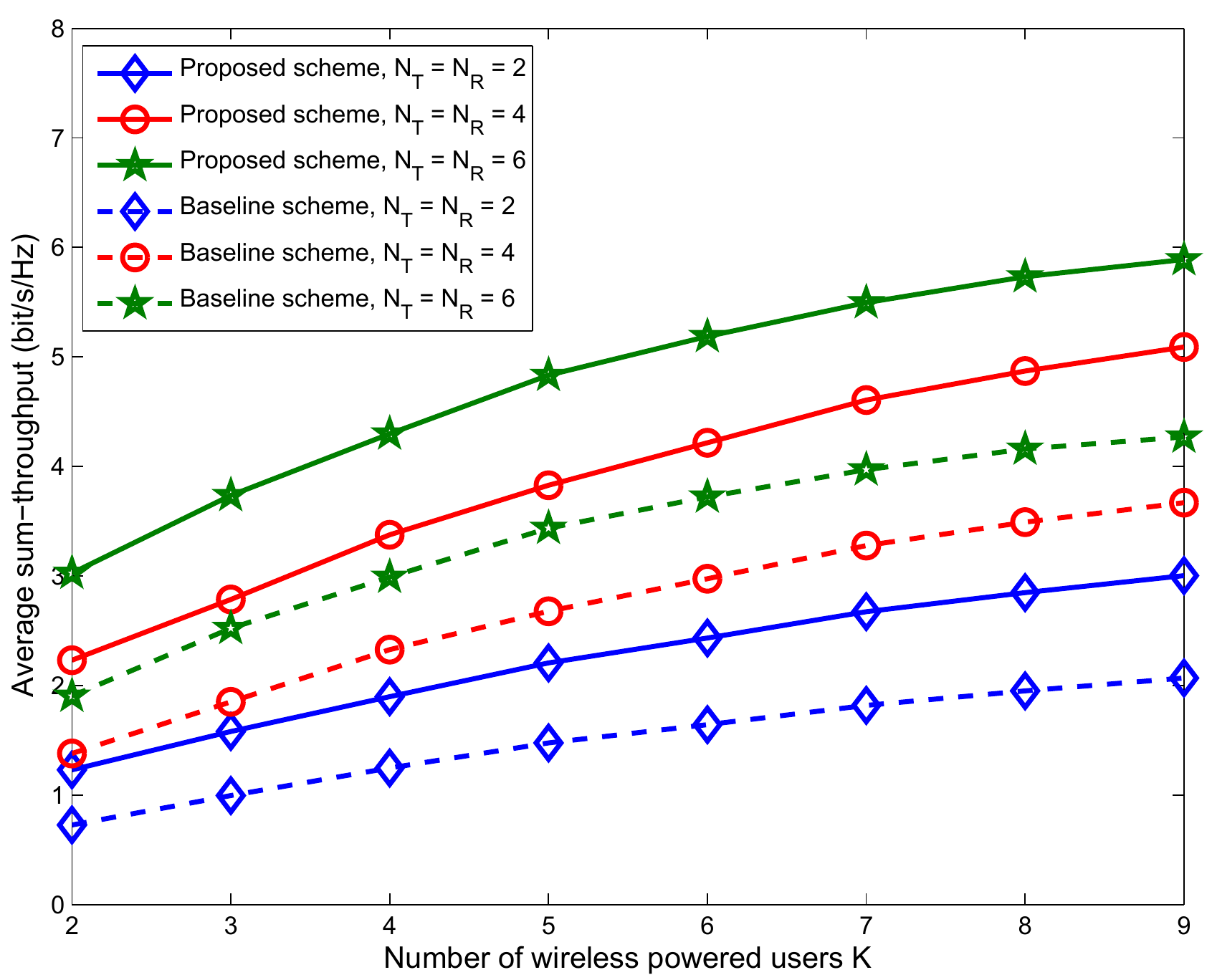}
\captionsetup{labelformat=empty}
\caption[]{(a)}\label{fig_2a}
\end{minipage}%
\begin{minipage}{0.5\linewidth}\ContinuedFloat
\centering
\includegraphics[width=1\linewidth]{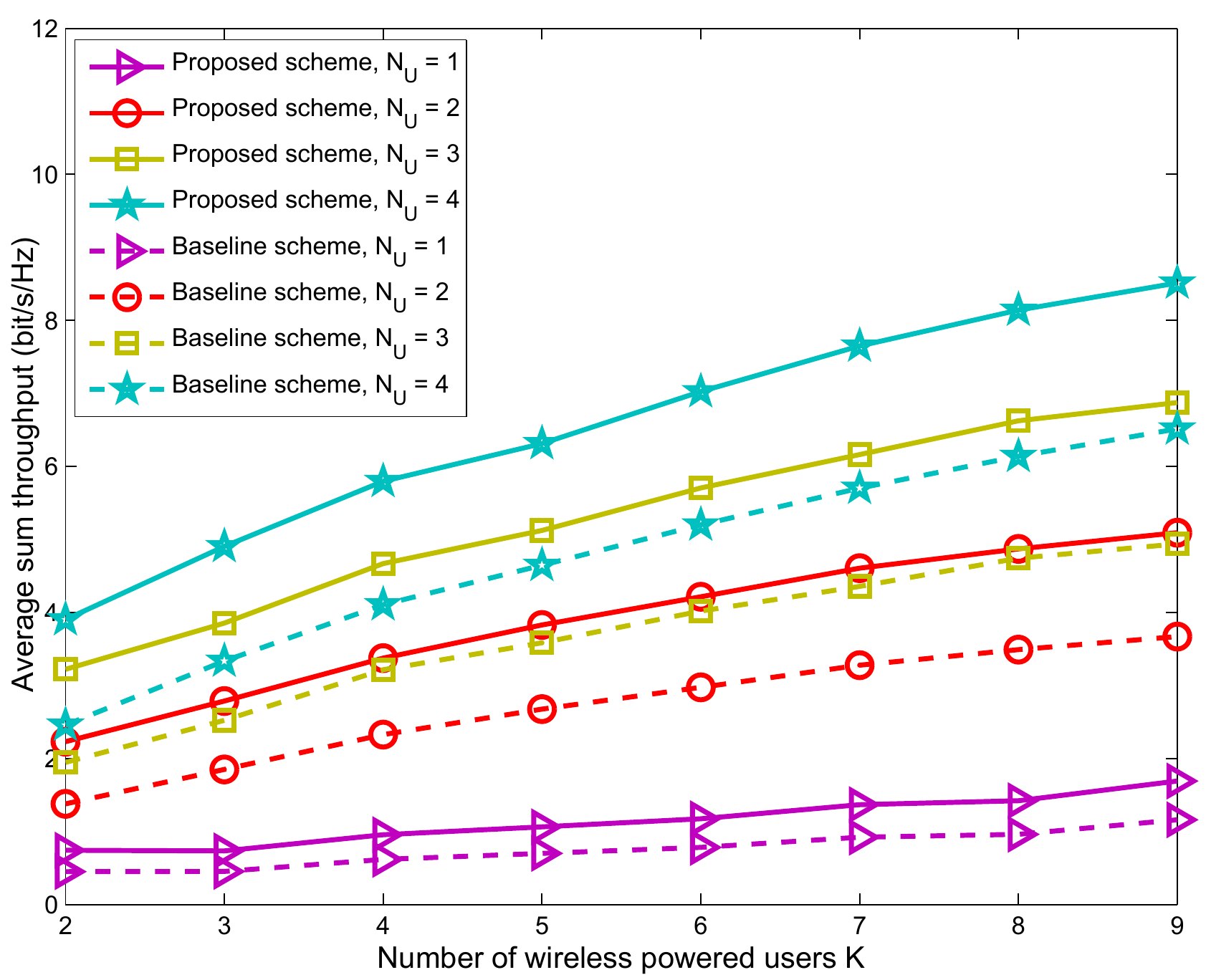}
\captionsetup{labelformat=empty}
\caption[]{(b)}\label{fig_2b}
\end{minipage}
\ContinuedFloat\caption{Average sum throughput (bit/s/Hz) versus numbers of wireless powered users $K$ for (a) different numbers of antennas at the power station and the information receiving station and (b) different numbers of antennas at the wireless powered users.}
\label{fig_2:av_sum_throughput_vs_}\vspace*{-4mm}
\end{figure}

In Figure \ref{fig_2a} (a), we show the average sum throughput versus the number of wireless powered users $K$  for different numbers of antennas equipped at the power station and the information receiving station, when the max-sum resource allocation scheme is employed. The value for the maximum transmit power at the power station is set to $P_{\max } = 35 $ dBm and the normalized maximum channel estimation error is $\sigma_{\mathrm{est}}^2 = 5\%$. With more wireless powered users in the system, the probability that there are users with good channel conditions is higher and performance improves, an effect known as multiuser diversity. In fact, the proposed max-sum resource allocation algorithm exploits multiuser diversity as it prefers to schedule users with good channel conditions to improve the system sum throughput. Additionally, the proposed scheme provides a substantial performance gain compared to the
baseline scheme, due to the resource allocation mismatch occurring for the latter scheme. 
The smaller amount of power harvested if the resource allocation is based on the (mismatched) linear EH model makes the wireless powered users more constrained in the uplink WIT period which reduces their contribution to the sum throughput. Figure \ref{fig_2a} (a) also shows the expected increase in sum throughput performance when the number of transmit antennas equipped at the power station and the number of receive antennas equipped at the information receiving station are increased. This is because the extra degrees of freedom offered by additional antennas enable an improved resource utilization. For instance,  with more transmit antennas equipped at the power station, the energy beam transmitted by the power station can be more efficiently steered in the direction of the wireless powered users which improves the efficiency of the WET. 
In Figure \ref{fig_2a} (b), we evaluate the average sum throughput versus the number of wireless powered users $K$, for different numbers of antennas equipped at the wireless powered users. We observe that for both the proposed and the baseline scheme there is a substantial gain in the sum throughput performance when the wireless powered users employ multiple antennas. The reason for this is twofold.
First, the additional antennas of the wireless powered users  act as additional wireless energy collectors which increase the total amount of harvested energy available for WIT. Second,  the additional user antennas increase the spatial multiplexing gain in the considered MIMO-WPCN. With multiple spatial data streams, there are more degrees of freedom available enabling a more flexible and more efficient resource allocation.
\begin{figure}\vspace*{-5mm}
\centering
\begin{minipage}{0.5\linewidth}
\centering
\includegraphics[width=1\linewidth]{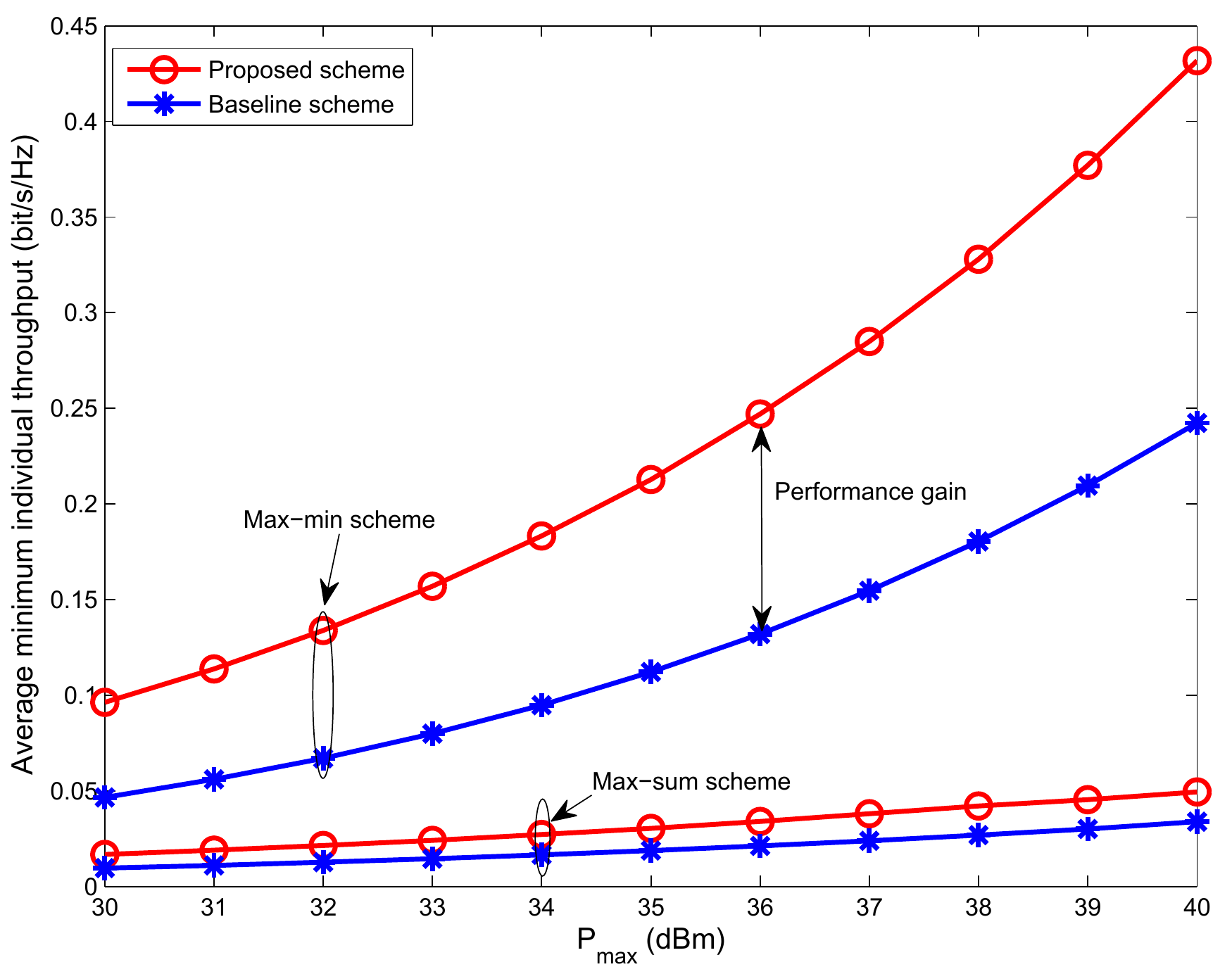}
\captionsetup{labelformat=empty}
\caption[]{(a)}\label{fig_3a}
\end{minipage}%
\begin{minipage}{0.5\linewidth}\ContinuedFloat
\centering
\includegraphics[width=1\linewidth]{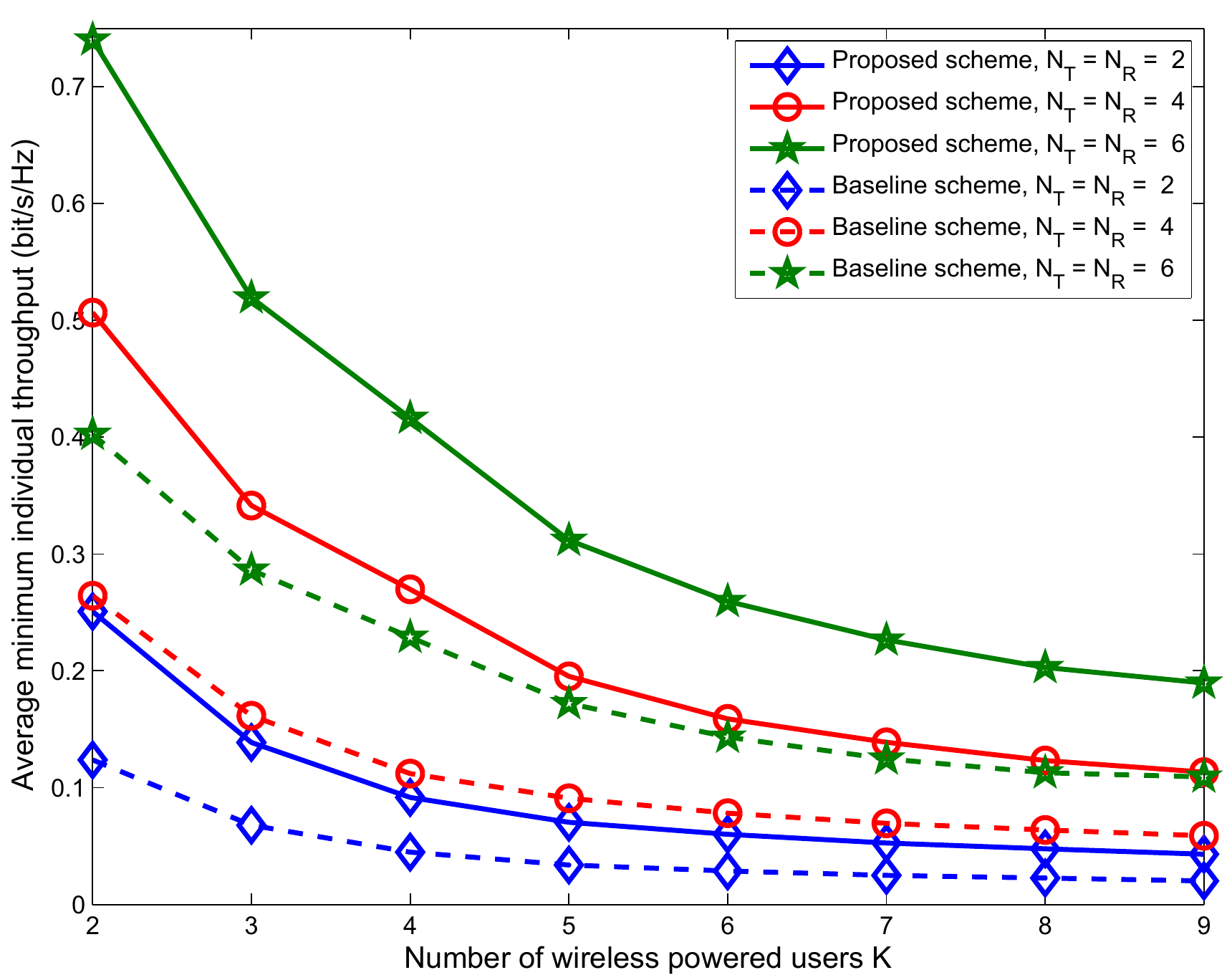}
\captionsetup{labelformat=empty}
\caption[]{(b)}\label{fig_3b}
\end{minipage}
\ContinuedFloat\caption{Average minimum individual throughput (bit/s/Hz) versus (a) the maximum transmit power at the power station $P_{\max }$ and (b) the number of wireless powered users $K$.}\vspace*{-4mm}
\label{fig_3:av_min_throughput_vs_}
\end{figure}

The average minimum individual throughput versus the maximum transmitted power $P_{\mathrm{max}}$ for $K = 4$ wireless powered users is depicted in Figure \ref{fig_3a} (a). As expected, the minimum individual throughput achieved with the max-min resource allocation scheme increases with the transmit power at the power station $P_{\max }$. On the other hand, the max-sum resource allocation scheme allocates most of the resources to the wireless powered users having the best channel conditions, and wireless powered users with relatively poor channel conditions are not allocated sufficient system resources. Hence, for the max-sum resource allocation scheme, the average minimum individual throughput increases only slowly with respect to the maximum transmit power $P_{\max }$.
Besides, for the max-min optimization, the performance gain of the proposed scheme over the baseline scheme is evident. On the contrary, the gain is significantly smaller for the max-sum scheme. In Figure \ref{fig_3b} (b), we show the average minimum individual throughput versus the number of wireless powered users $K$, for different numbers of antennas equipped at the power station and the information receiving station. As the number of wireless powered users increases, the constraint on the minimum individual throughput becomes more stringent, cf. C8 in \eqref{eqn:updated_problem_final_max_min}, since the resource allocator is required to ensure fairness to a larger number of wireless powered users, despite their potentially poor channel conditions. Thus, in contrast to the sum throughput, cf. Figure \ref{fig_2a} (a), the average minimum individual throughput decreases with increasing $K$. This comparison demonstrates that there is a non-trivial trade-off between fairness and the overall maximum sum throughput for multiuser MIMO-WPCNs. However, the performance degradation in terms of the average minimum individual throughput is alleviated when the power station and the information receiving station are both equipped with multiple antennas, due to the increased degrees of freedom for resource allocation.\\
\begin{figure}
\begin{minipage}[t]{0.47\linewidth}
\centering
\includegraphics[width = 3.15 in]{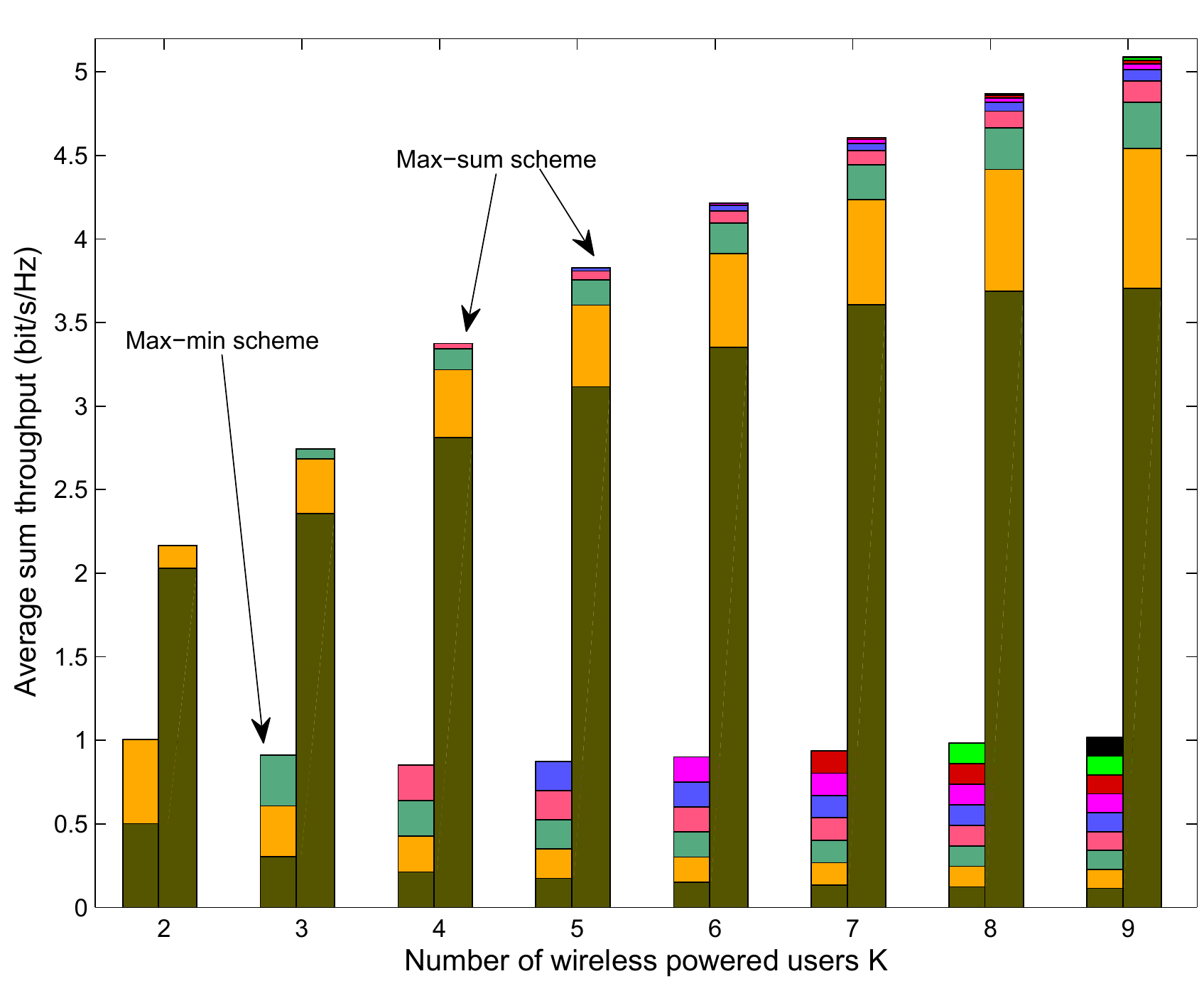}
\caption[width=2 in]{Average sum throughput (bit/s/Hz) versus the number of wireless powered users $K$ for $P_{\max } = 35$ dBm and $\sigma_{\text{est}}^2 = 0.05$.}\vspace*{-4mm}
\label{fig_3:av_sum_fairness}
\end{minipage}
\hspace{0.3cm}
\begin{minipage}[t]{0.47\linewidth}
\centering
\includegraphics[width=3.2 in]{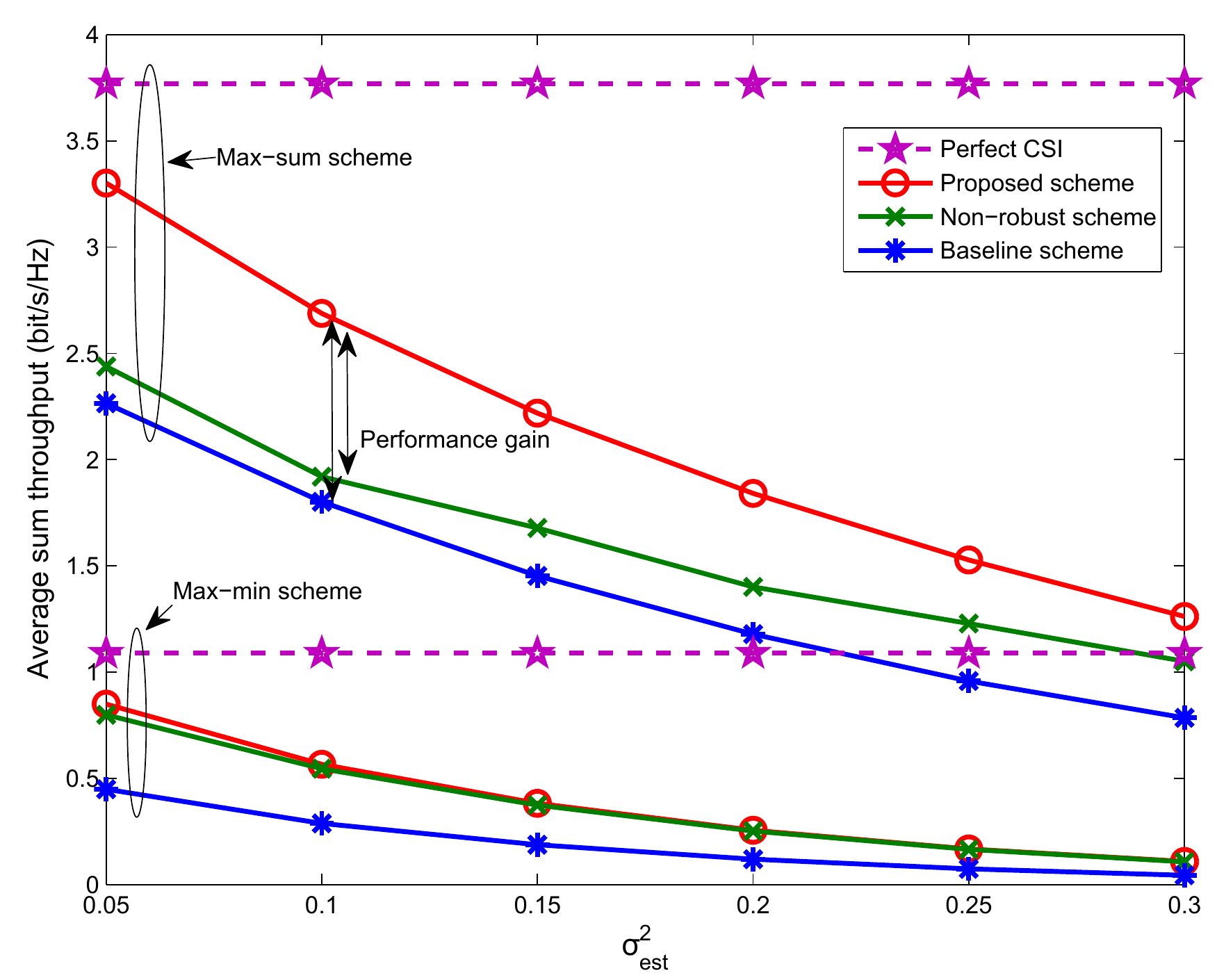}
\caption[width=2 in]{Average sum throughput (bit/s/Hz) versus normalized maximum channel estimation error $\sigma_{\mathrm{est}}^2$ for $K=4$ wireless powered users and $P_{\mathrm{max}} = 35 $ dBm.}
\label{fig_4:av_min_throughput_vs_sigma_4_users}\vspace*{-4mm}
\end{minipage}
\end{figure}
To further investigate the non-trivial trade-off between the sum throughput and resource allocation fairness for the wireless powered users, in Figure \ref{fig_3:av_sum_fairness}, we depict the average sum throughput versus the number of wireless powered users. For a given number of users $K$, the bars on the left show the rate allocation of the users for the proposed max-min scheme, while the bars on the right illustrate the rate allocation of the users for the proposed max-sum scheme. Different colors are used for the throughputs of different wireless powered users. As can be observed, the max-sum scheme allocates the resources in an unfair manner. In particular, the user with the best channel conditions consumes most of the system resources as this maximizes the sum throughput performance. For the max-min scheme, on the other hand, the available resources are allocated in such a manner that all $K$ wireless powered users achieve identical throughputs in order to guarantee fairness. In fact, for the max-min scheme, the throughput is limited by the user with the worst channel condition, since the resources are allocated such that the individual throughput achieved by each wireless powered user is equalized as much as possible. As the number of wireless powered users increases, the constraints for optimization become more stringent since the same system resources have to be shared by more users and the probability of a user with poor channel conditions increases. This leads to a limited sum throughput performance for the max-min scheme compared to the increase in sum throughput for the max-sum scheme.

Figure \ref{fig_4:av_min_throughput_vs_sigma_4_users} shows the average sum throughput versus the normalized maximum channel estimation error $\sigma_{\mathrm{est}}^2$ for $K=4$ wireless powered users and a maximum transmit power allowance of $P_{\mathrm{max}} = 35 $ dBm. As can be observed, the average sum throughput performance degrades for both the max-sum and the max-min resource allocation schemes as the quality of the CSI decreases, since the spatial degrees of freedom and the time resources cannot be efficiently exploited for resource allocation anymore. Additionally, as the normalized maximum channel estimation error $\sigma_{\mathrm{est}}^2$ increases, the performance of the proposed schemes approaches that of the baseline scheme because optimal resource allocation cannot bypass the large estimation errors dominating the overall performance. In Figure \ref{fig_4:av_min_throughput_vs_sigma_4_users}, we also show the average sum throughput performance of a non-robust scheme. The considered non-robust scheme is optimized for the non-linear EH model but treats the estimated CSI of the channel between the power station and the wireless powered users as perfect CSI for resource allocation. This scheme still takes into account the imperfections of the CSI estimation of the channel between the wireless powered users and the information receiving station. The non-robust scheme fails to allocate sufficient energy resources in the WET period. As a result, the wireless powered users are unable to fully utilize the space and time resources needed for maximizing their performance in the WIT period. 
Hence, the non-robust scheme results in a lower sum throughput performance compared to the proposed resource allocation scheme, especially for max-sum resource allocation. Lastly, in Figure \ref{fig_4:av_min_throughput_vs_sigma_4_users}, we have included simulation results for a benchmark scheme based on the non-linear EH model and perfect CSI knowledge. The scheme having access to perfect CSI can fully utilize the degrees of freedom offered by the multiple antennas for resource allocation. Thus, the average sum throughput achieved by this scheme serves as a performance upper bound for the other schemes that have access to  imperfect CSI only.
\begin{figure}\vspace*{-5mm}
\centering
\begin{minipage}{0.5\linewidth}
\centering
\includegraphics[width=1\linewidth]{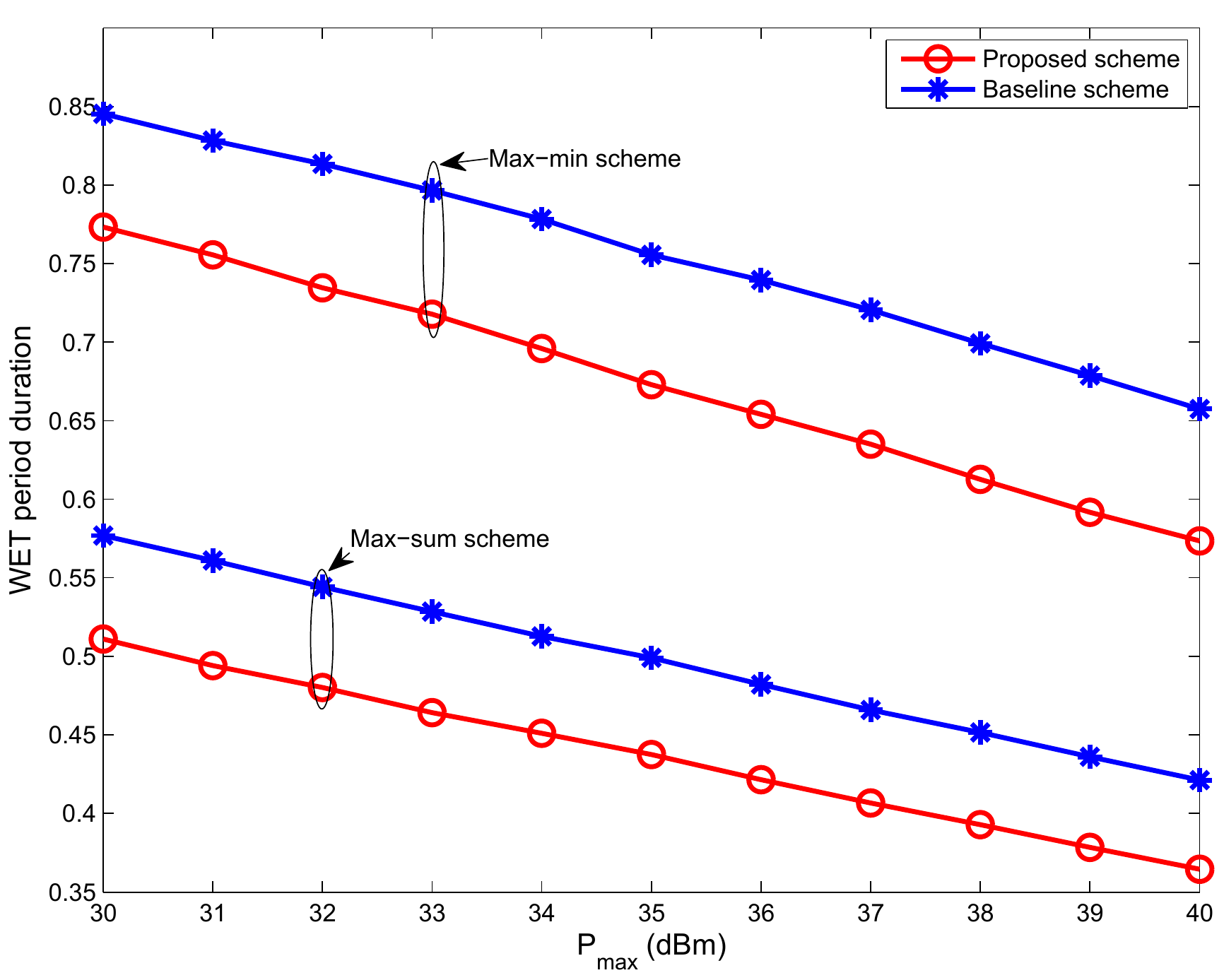}
\captionsetup{labelformat=empty}
\caption[]{(a)}\label{fig_9a}
\end{minipage}%
\begin{minipage}{0.5\linewidth}\ContinuedFloat
\centering
\includegraphics[width=1\linewidth]{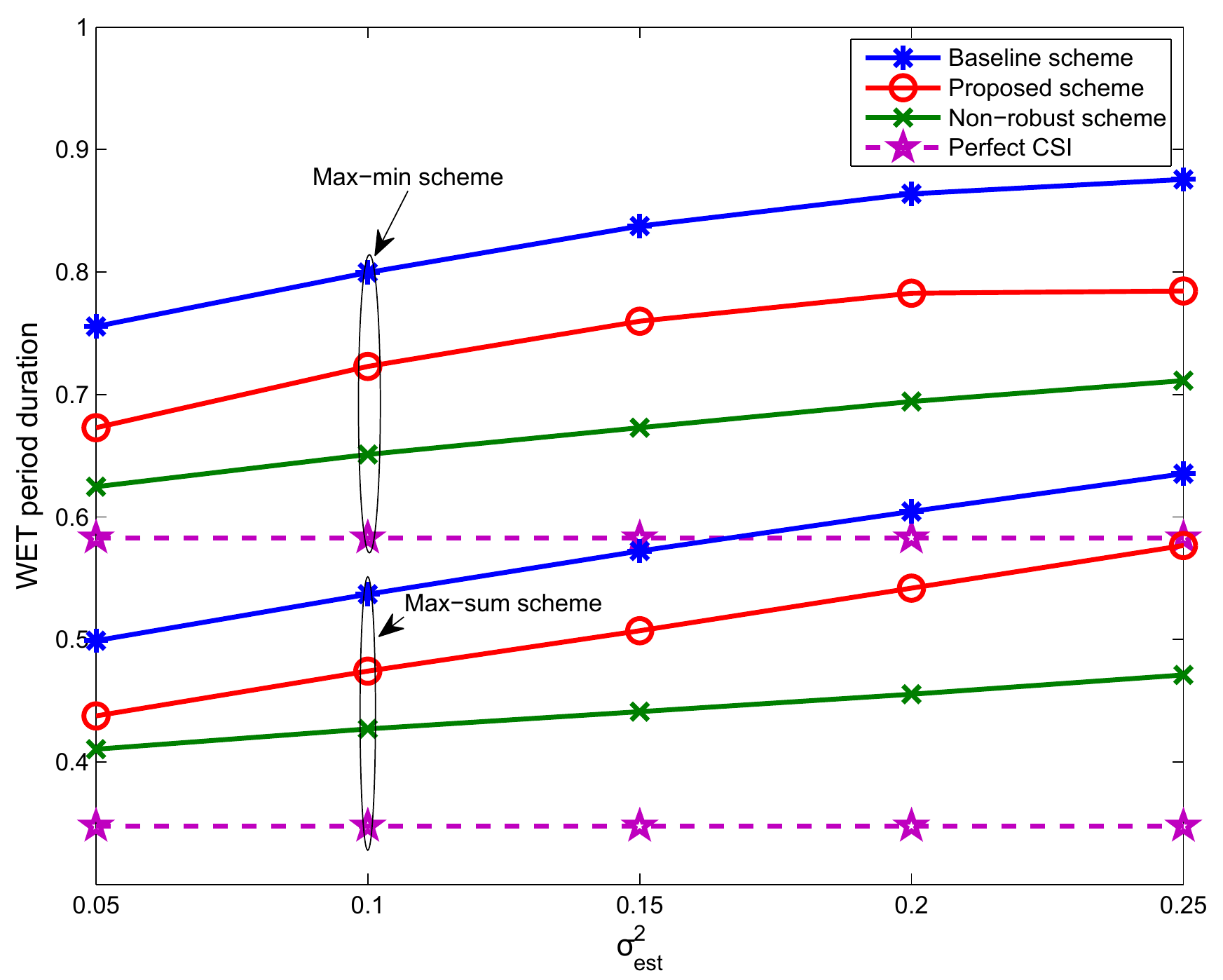}
\captionsetup{labelformat=empty}
\caption[]{(b)}\label{fig_9b}
\end{minipage}
\ContinuedFloat\caption{WET period duration $\tau_0$ versus (a) the maximum transmit power at the power station $P_{\text{max}}$ and (b) the normalized maximum channel estimation error $\sigma_{\mathrm{est}}^2$ for $K=4$ wireless powered users.}
\label{fig_resps:WET_period_duration}\vspace*{-4mm}
\end{figure}

Figure \ref{fig_resps:WET_period_duration} (a) depicts the duration of the WET period $\tau_0$ versus the maximum transmit power at the power station $P_{\text{max}}$ for $\sigma_{\text{est}}^2 = 0.05$ and $K=4$ wireless powered users.  The duration of the WET period is significantly larger for the baseline scheme, for both the max-sum and max-min schemes, compared to the proposed resource allocation schemes. Moreover, the WET period duration is larger for the max-min resource allocation schemes, compared to the max-sum schemes. This is because wireless powered users with poor channel conditions need to harvest more energy in the downlink WET phase in order to achieve the required minimum individual throughput for uplink WIT. Moreover, as the transmit power increases, the wireless powered users require less time to harvest the energy required for information transmission in the WIT period. In Figure \ref{fig_resps:WET_period_duration} (b), we show the duration of the WET period $\tau_0$ versus the normalized maximum channel estimation error $\sigma_{\text{est}}^2$ for $P_{\text{max}} = 35$ dBm and $K=4$ wireless powered users for all of the considered schemes. As the normalized maximum channel estimation error $\sigma_{\text{est}}^2$ increases, both the proposed and the baseline resource allocation schemes require a longer time for WET in order to ensure that the users harvest the amount of energy required for maximization of the system design objectives. In fact, with increasing CSI uncertainty, the energy beam cannot be accurately steered towards the wireless powered users for EH, and a longer time for WET is needed in order to ensure constraint C3 is satisfied. This observation coincides with our previous conclusions from the analytical solution in \eqref{eqn:tau0prop}. On the contrary, the WET period duration for the non-robust scheme does not increase significantly with increasing channel estimation error variance. This is because the non-robust scheme is not optimized for combating the CSI estimation errors of the channel between the power station and the wireless powered users and it does not modify the resource allocation in the WET period in case of higher CSI uncertainty. A constant WET period duration can also be observed when perfect CSI is available for resource allocation, i.e., for the benchmark scheme, since the perfect CSI scheme does not suffer from channel estimation errors. Considering the proposed, the baseline, and the non-robust schemes, it can be observed in Figure \ref{fig_resps:WET_period_duration} (b) that for large CSI estimation error variances, the optimal WET period duration $\tau_0^*$ starts to saturate. The reason for this saturation is that a further increase in the WET period duration will not lead to an optimal resource allocation policy with respect to the system objectives, since the WIT duration has to decrease accordingly, cf. constraint C2. If the WIT duration is too short, the wireless powered users will not be able to achieve a high throughput even if they have harvested a large amount of energy because of the long WET period. Hence, the one-dimensional search for the optimal WET period duration always results in a $\tau_0$ that strikes a balance between the time for WET and the time for WIT.
\section{Conclusions}\label{sect:conclusion}
In this paper, we studied robust resource allocation schemes for MIMO-WPCNs based on a practical non-linear EH model. In particular, we designed joint time allocation and power control resource allocation algorithms for maximizing the total system throughput and for maximizing the minimum individual throughput at the wireless powered users, respectively. The resource allocation algorithm designs were formulated as non-convex optimization problems, which were efficiently solved by utilizing a one-dimensional search, where in each iteration a convex optimization problem was solved. Simulation results demonstrated that the proposed resource allocation schemes achieve significant gains in throughput compared to baseline resource allocation schemes optimized for the traditional linear EH model. Besides, the obtained results unveiled the trade-off between achieving maximum system throughput and ensuring fairness among the wireless powered users in MIMO-WPCNs. Finally, the developed resource allocation schemes were shown to be robust against imperfect CSI knowledge.
\section*{Appendix}
\subsection{Proof of Theorem \ref{sdp_relaxation}}\label{rank_one_proof}
We follow a similar approach as in \cite{JR:sum_rate_mimo_WPCN_rank_one} to prove Theorem \ref{sdp_relaxation}. We aim to show the structure of the optimal beamforming matrix $\mathbf{V}^*$. To this end,  we consider the KKT conditions of \eqref{eqn:updated_problem_final}, cf. \eqref{KKT_conditions}--\eqref{subseq:eq6}. Since the columns of $\mathbf{V}^*$ lie in the null space of $\mathbf{M}_{\mathrm{C5}}^*$, cf. \eqref{subseq:eq3}, we study the rank and null space of $\mathbf{M}_{\mathrm{C5}}^*$ for obtaining the structure of $\mathbf{V}^*$. Thus, exploiting \eqref{subseq:eq6}, we have the following equation:\vspace*{-2mm}
\begin{equation}\label{eqn:derivative_v}
\mathbf{M}_{\mathrm{C5}}^*   = \mu^* \mathbf{I}_{N_{\mathrm{T}}}  - \sum_{k=1}^{K} \sum_{l=1}^{N_{\tiny \mathrm{U}_k} }\Big[\mathbf{U}_{\widehat{\mathbf{g}}_k} \mathbf{M}_{\mathrm{C3b}_k}^*\mathbf{U}_{\widehat{\mathbf{g}}_k}^H\Big]_{a:b,c:d},
\end{equation}
where $a = (l-1)N_{\mathrm{T}} + 1, b=l N_{\mathrm{T}}, c= (l-1)N_{\mathrm{T}} + 1, $ $d=l N_{\mathrm{T}}$. For notational simplicity, we define $\mathbf{\Gamma} = \sum_{k=1}^{K} \sum_{l=1}^{N_{\tiny \mathrm{U}_k} }\Big[\mathbf{U}_{\widehat{\mathbf{g}}_k} \mathbf{M}_{\mathrm{C3b}_k}^*\mathbf{U}_{\widehat{\mathbf{g}}_k}^H\Big]_{a:b,c:d}\succeq \zero$ which is a Hermitian matrix. From \eqref{subseq:eq1}, since matrix $\mathbf{M}_{\mathrm{C5}}^* = \mu^* \mathbf{I}_{N_\mathrm{T} } - \mathbf{\Gamma}$ is positive semi-definite,
\vspace*{-2mm}\begin{eqnarray}
\mu^* \geq \lambda_{\mathbf{\Gamma}}^{\max} &\ge& 0,
\end{eqnarray}
must hold, where $\lambda_{\mathbf{\Gamma}}^{\max}$ is the a real-valued maximum eigenvalue of matrix $\mathbf{\Gamma}$. Considering the KKT condition related to matrix $\mathbf{V}^*$ in \eqref{subseq:eq3}, we can show that if $\mu^* > \lambda_{\mathbf{\Gamma}}^{\max} $, matrix $\mathbf{M}_{\mathrm{C5}}^*$ will become positive definite and full rank. However, this will yield the solution $\mathbf{V}^* = \zero$ which contradicts the KKT condition in \eqref{subseq:eq1_1} as $\mu^* > 0$ and $P_{\max}>0$. Thus, for the optimal solution, the dual variable $\mu^*$ has to be equal to the largest eigenvalue of matrix $\mathbf{\Gamma}, $ i.e., $\mu^* = \lambda_{\mathbf{\Gamma}}^{\max}$. Besides, in order to have a bounded optimal dual solution, it follows that the null space of $\mathbf{M}_{\mathrm{C5}}^*$ is spanned by vector $\mathbf{u}_{\mathbf{\Gamma},\max}$, which is the unit-norm eigenvector of $\mathbf{\Gamma}$ associated with eigenvalue $\lambda_{\mathbf{\Gamma}}^{\max}$. As a result,  we obtain the structure of the optimal energy matrix $\mathbf{V}^*$ as
\begin{equation}
\mathbf{V}^* = \delta \mathbf{u}_{\mathbf{\Gamma},\max} \mathbf{u}_{\mathbf{\Gamma},\max}^H.
\end{equation}
Additionally, since $\mu^* > 0$ and $P_{\max}>0$, we conclude that $\delta = P_{\max }$ and $\Tr(\mathbf{V}^*) = P_{\mathrm{max}}$ holds at the optimal solution. \qed
\subsection{Proof of Proposition 1}\label{time_allocation_prop}
For problem \eqref{eqn:updated_problem_final}, it can be easily shown that for the optimum solution, $\sum_{k=0}^{K}\tau_k = T_{\max}$ holds. Hence, it follows from \eqref{subseq:eq1_2} that $\kappa^* > 0 $. Additionally, we can show that $T_{\max}P_{\mathrm{c}_k} + \sum_{i=1}^{\mathrm{min}\{N_{\tiny \mathrm{U}_k},N_{\text{\tiny{R}}}\}}\widetilde{\lambda}_{i, k}^* \varepsilon_k = \tau_0^* \Phi_k^{\mathrm{Practical}}(\theta_k^*), \forall k$, i.e., $\beta_k^* > 0, \forall k$, must hold.
Based on the previous observations, we obtain
\begin{equation}\label{app:eq7}
\sum_{i=1}^{\mathrm{min}\{N_{\tiny \mathrm{U}_k},N_{\text{\tiny{R}}}\}}\widetilde{\lambda}_{i, k}^* = \frac{\tau_0 \Phi_k^{\mathrm{Practical}}(\theta_k^*) - T_{\max}P_{\mathrm{c}_k}}{\varepsilon_k}, \forall i, k.
\end{equation}
From \eqref{app:eq7}, we can express $\tau_{k}^*$ as:
\begin{equation}\label{tau_k_optimal}
\tau_k^*  = \frac{\tau_0 \Phi_k^{\mathrm{Practical}}(\theta_k^*) - T_{\max}P_{\mathrm{c}_k}}{\sum_{i=1}^{\mathrm{min}\{N_{\tiny \mathrm{U}_k},N_{\text{\tiny{R}}}\}}\lambda_{i, k}^* \varepsilon_k}, \forall k.
\end{equation}
Combining \eqref{tau_k_optimal} and $\tau_0^* + \sum_{k=1}^{K}\tau_k^* = T_{\max}$ yields
\begin{eqnarray}\label{tau_0_optimal}
\tau_0^*\Big(1+\sum_{k=1}^{K}\frac{\Phi_k^{\mathrm{Practical}}}{\sum_{i=1}^{\mathrm{min}\{N_{\tiny \mathrm{U}_k},N_{\text{\tiny{R}}}\}}\lambda_{i, k}^* \varepsilon_k}\Big) - \sum_{k=1}^{K}\frac{T_{\max}P_{\mathrm{c}_k}}{\sum_{i=1}^{\min\{N_{\tiny \mathrm{U}_k},N_{\text{\tiny{R}}}\}}\lambda_{i, k}^* \varepsilon_k} = T_{\max},
\end{eqnarray}
which leads to the final equation \eqref{eqn:tau0prop} and concludes the proof. \qed

\vspace*{-5mm}

\bibliographystyle{IEEEtran}
\bibliography{vicky_leng}

\end{document}